\documentclass[aps,pra,twocolumn,a4paper,floatfix,
superscriptaddress, nofootinbib]{revtex4-2}
  \usepackage[T1]{fontenc}
  \usepackage{amsmath}
  \usepackage{amsthm}
  \usepackage{amsbsy}
  \usepackage{amssymb}
  \usepackage{enumitem}
\usepackage{lipsum}
\usepackage[colorlinks=true,urlcolor=blue,citecolor=blue,linkcolor=blue]{hyperref}
\usepackage{graphicx, color}
\usepackage[dvipsnames,table,xcdraw]{xcolor}

\usepackage{orcidlink}

\newcommand{\D}{{\rm{d}}}

\DeclareMathOperator*{\argmax}{arg\,max}

\newcommand{\ket}[1]{\left|#1\right\rangle}
\newcommand{\bra}[1]{\left\langle#1\right|}

\DeclareMathOperator{\Ext}{Ext}
\DeclareMathOperator{\Tr}{Tr}
\DeclareMathOperator{\Conv}{Conv}
\newtheorem{statement}{Statement}
\newtheorem*{statement*}{Statement}
\newtheorem{definition}{Definition}

\begin{document}
\title{Tight inequalities for nonclassicality of measurement statistics}

\author{V. S. Kovtoniuk \orcidlink{0009-0008-8363-0898}}
\affiliation{Bogolyubov Institute for Theoretical Physics, NAS of Ukraine, Vulytsya Metrolohichna 14-b, 03143 Kyiv, Ukraine}

\author{E. V. Stolyarov \orcidlink{0000-0001-7531-5953}}
\affiliation{Bogolyubov Institute for Theoretical Physics, NAS of Ukraine, Vulytsya Metrolohichna 14-b, 03143 Kyiv, Ukraine}
\affiliation{Institute of Physics, NAS of Ukraine, Prospect Nauky 46, 03028 Kyiv, Ukraine}

\author{O. V. Kliushnichenko \orcidlink{0000-0003-4806-8971}}
\affiliation{Institute of Physics, NAS of Ukraine, Prospect Nauky 46, 03028 Kyiv, Ukraine}

\author{A. A. Semenov \orcidlink{0000-0001-5104-6445}}
\affiliation{Bogolyubov Institute for Theoretical Physics, NAS of Ukraine, Vulytsya Metrolohichna 14-b, 03143 Kyiv, Ukraine}
\affiliation{Institute of Physics, NAS of Ukraine, Prospect Nauky 46, 03028 Kyiv, Ukraine}
\affiliation{Kyiv Academic University, Boulevard Vernadskogo  36, 03142  Kyiv, Ukraine}

\begin{abstract}
In quantum optics, measurement statistics---for example, photocounting statistics---are considered nonclassical if they cannot be reproduced with  statistical mixtures of classical radiation fields.
We have formulated a necessary and sufficient condition for such nonclassicality.
This condition is  given by a set of inequalities that tightly bound the convex set of probabilities  associated with classical electromagnetic radiation.
Analytical forms for full sets and subsets of these inequalities are obtained for  important cases of realistic photocounting measurements and unbalanced homodyne detection.
As an example, we consider photocounting statistics of phase-squeezed coherent states.
Contrary to a common intuition, the analysis developed here reveals distinct nonclassical properties of these statistics that can be experimentally corroborated with minimal resources.
\end{abstract}

\maketitle

\section{Introduction}

	The concept of nonclassical states in quantum optics \cite{titulaer65,mandel86,mandel_book,vogel_book,agarwal_book,Schnabel2017,sperling2018a,sperling2018b,sperling2020} is based on the fact that all measurements under the coherent states $\ket{\alpha}$ of a light mode can be explained with no need to quantize the electromagnetic field.
	Since the density operator $\hat{\rho}$ of any quantum state can be represented as
		\begin{align}\label{Eq:Pfunction}
				\hat\rho=\int_\mathbb{C}d^2 \alpha P(\alpha) \ket{\alpha}\bra{\alpha},
			\end{align}
	where $P(\alpha)$ is the Glauber-Sudarshan $P$ function \cite{glauber63c,sudarshan63}, the states with $P(\alpha)\geq0$ can be regarded as classical statistical mixtures of coherent states.
	The remaining states are considered to be nonclassical.

	Determining whether a given quantum state is nonclassical can be a challenging task, since the $P$ function is often not regular.
	A wide range of nonclassicality tests focuses on the complete (as far as achievable in experiments) information about quantum states \cite{kiesel08,kiesel11b,kiesel10,Agudelo2013,luis15,bohmann2020,bohmann2020b,Hillery1987,Lee1991,Asboth2005, Miranowicz2015a,Miranowicz2015b,Yadin2019,Luo2019}.
	Another group of nonclassicality tests involves partial information obtained through specific measurements or sets of measurements; see, e.g., Refs.~\cite{Stoler1970, Stoler1971,mandel79,reid1986,Wu1986,Wu1987, Vahlbruch, agarwal92,agarwal93,klyshko1996,vogel00,richter02,rivas2009,sperling12c,bartley13,sperling13b,park2015a,park2015b,richter02,Perina2020,Semenov2021,Innocenti2023}.

	The latter tests provide only sufficient conditions for nonclassicality of quantum states.
	However, they address a more specific but still crucial issue: can the given measurement statistics be reproduced with a statistical mixture of coherent states?
	An example of such a measurement statistic is given by the photocounting distribution $\mathcal{P}(n)$ estimated with photon-number resolving (PNR) detectors, where $n$ is the number of photons.
	The corresponding archetypal tests are based on the Mandel $Q$ parameter \cite{mandel79}, its higher-order generalization \cite{agarwal92}, the Klyshko criterion \cite{klyshko1996}, etc.
	The first two have been adapted \cite{sperling12c,bartley13,sperling13b} to a realistic scenario with click detectors \cite{paul1996,castelletto2007,schettini2007,blanchet08,achilles03,fitch03,rehacek03,sperling12a}.
	Other examples of nonclassicality tests \cite{park2015a,park2015b} relevant to this paper involve a different measurement procedure---unbalanced homodyne detection \cite{wallentowitz96}.
	In this case, we deal with a set of photocounting distributions $\mathcal{P}(n|\gamma)$ conditioned on a device setting---the complex amplitude $\gamma$ of the local oscillator.

	Nonclassicality of a given measurement statistic implies that the set of probability distributions $\mathcal{P}(A|a)$ to obtain the value $A$ under a measurement with the setting $a$ cannot be reproduced with a statistical mixture of coherent states.
	This implies that $\mathcal{P}(A|a)$ cannot be expressed as a convex combination of the probability distributions $\Pi(A|a;\alpha)=\bra{\alpha}\hat{\Pi}(A|a)\ket{\alpha}$ for the coherent states $\ket{\alpha}$.
    Here $\hat{\Pi}(A|a)$ is an element of the positive operator-valued measure (POVM), corresponding to the measurement outcome $A$ with the setting $a$.
    The functions $\Pi(A|a;\alpha)$ \cite{cahill69,cahill69a} are also referred to as $Q$ symbols of the POVM.
	If $A$ is the number of photons and no detector settings are involved, nonclassicality of measurements statistics aligns with nonclassicality of photon-number statistics in Ref.~\cite{Innocenti2022}.
	Similar to Ref.~\cite{Innocenti2023}, our approach extends this concept to a set of observables, each of them is assigned an associated value of the device setting $a$.

	As it has been shown in Refs.~\cite{Semenov2021,Innocenti2022,Innocenti2023,Kovtoniuk2022} (see also Ref.~\cite{rivas2009} for particular cases),
	the probability distributions $\mathcal{P}(A|a)$ are nonclassical if and only if there exists a function $\lambda(A, a)$, which we refer to as the test function, that the inequality
	\begin{align}\label{Eq:BellLikeInequalities}
		\sum_{a,A} \mathcal{P}(A|a) \lambda(A, a) \leq \sup_{\alpha \in \mathbb{C}} \sum_{a,A} \Pi(A|a;\alpha) \lambda(A, a)
	\end{align}
	is violated; cf. Appendix~\ref{App:DDBell} for details.
	The left-hand side of this inequality can be estimated from the measured data, facilitating practical implementation of the test.
    On the right-hand side, the supremum over the coherent amplitude $\alpha$ in the entire complex plane $\mathbb{C}$ is taken.

    Inequality~(\ref{Eq:BellLikeInequalities}) implies that the sum of expectation values of the test function $\lambda(A, a)$ over all settings $a$ for the classical measurement statistics does not exceed the same sum of its expectation values for all coherent states $\ket{\alpha}$.
    This inequality can also be considered as a consequence of the supporting hyperplane theorem \cite{boyd_book}.
    A similar concept is used for testing other quantum phenomena.
    For example, a generalized form of Bell inequalities and entanglement witness are based on the same mathematical idea of separating the convex sets of local realistic probability distributions and separable quantum states by supporting hyperplanes; see, e.g., Refs.~\cite{Brunner2014,Semenov2021,Kovtoniuk2022,horodecki96,terhal99,horodecki09}.

	Finding the optimal test-function $\lambda(A, a)$ in inequality (\ref{Eq:BellLikeInequalities}) can be a challenging problem.
	In this paper, we find that the amount of these inequalities can be reduced to
    tightly bound the convex set of classical probability distributions.
	Thus, the problem of finding the optimal $\lambda(A, a)$ is drastically simplified to finding violations among tight inequalities.

	In particular, our technique enables one to verify
    nonclassicality of photocounting statistics for phase-squeezed coherent states using minimal experimental resources.
    The obtained result contradicts the intuitive expectation based on positive variance of excess noise for photocounting measurements with such states  \cite{Schnabel2017}, implying the positive Mandel $Q$ parameter.
    We also apply our method to unbalanced homodyne detection.
    In this context, we have demonstrated nonclassicality of the measurement statistics conditioned on the value of the local-oscillator amplitude and obtained with a single on-off detector for a squeezed vacuum state.
    Importantly, the local-oscillator phase is chosen to be the same as the phase of the antisqueezed quadrature, which also makes nonclassicality of this measurement statistics counterintuitive.

    The rest of the paper is organized as follows.
    In Sec.~\ref{Sec:Tight} we employ an analysis based on convex geometry to formulate a general framework for tight inequalities testing nonclassicality of measurement statistics.
    These inequalities are applied to realistic photocounting measurements in Sec.~\ref{Sec:PC}.
    In Sec.~\ref{Sec:UHD} they are derived for unbalanced homodyne detection.
    A summary and concluding remarks are given in Sec.~\ref{Sec:Concl}.

\section{Tight inequalities: geometrical framework}
\label{Sec:Tight}
	Our goal is to optimize the set of all test functions $\lambda(A,a)$, i.e., to construct the minimal set of $\lambda(A,a)$ for which violation of inequalities (\ref{Eq:BellLikeInequalities}) is the necessary and sufficient condition for nonclassicality of the probability distributions $\mathcal{P}(A|a)$.
	We achieve this by employing the methods of convex geometry; see, e.g., Refs.~\cite{Rockafellar_book, Hug_book}.
	The corresponding inequalities tightly bound the convex set of classical probability distributions and thus referred to as the tight inequalities.
	Formalizing the requirements of their completeness and minimality, we can give the following definition.
	\begin{definition}\label{DefTight}
		Let $\Lambda$ be a set of test functions $\lambda(A,a)$.
		We say that they produce tight inequalities if
		\begin{enumerate}[label=(\roman*)]
			\item\label{Def1-1} for any $\lambda{^{\prime}}(A,a) \not \in \Lambda$, the corresponding inequality can be derived from tight inequalities;
			\item\label{Def1-2} no tight inequality can be derived from other tight  inequalities.
		\end{enumerate}
	\end{definition}
    In the context of this definition, deriving an inequality from others involves two operations: both sides of inequalities can be multiplied by non-negative numbers, and different inequalities can be added to each other. 
Thus, the inequality with $\lambda(A,a)$ follows from the inequalities with $\lambda_i(A,a)$ if $\lambda(A,a)=\sum_{i}C_i\lambda_i(A,a)$ such that $C_i\geq 0$.
In other words, in such a case, $\lambda(A,a)$ is a conic combination of $\lambda_i(A,a)$.

    In what follows, we also use the vector notations $\boldsymbol{\mathcal{P}}$ and $\boldsymbol{\lambda}$ for the probability distributions $\mathcal{P}(A|a)$ and the test function $\lambda(A,a)$, respectively.
    First of all, we note that the values of the probability distributions $\mathcal{P}(A|a)$ are not independent due to the normalization conditions,
        \begin{align}\label{Eq:Normalization}
            \sum\limits_A\mathcal{P}(A|a)=1.
        \end{align}
    We define the vector $\boldsymbol{\mathcal{P}}$ as the set of elements corresponding to the independent values of these probability distributions.
    Hence, the dimensionality of the vector $\boldsymbol{\mathcal{P}}$ is given by
        \begin{align}\label{Eq:Dimensionality}
            \dim \boldsymbol{\mathcal{P}}=(N_{\mathrm{A}}-1)N_{\textrm{a}},
        \end{align}
    where $N_{\mathrm{A}}$ is the number of all measurement outcomes $A$ and $N_{\mathrm{a}}$ is the number of device settings $a$.
    Application of normalization condition (\ref{Eq:Normalization}) to inequality~(\ref{Eq:BellLikeInequalities}) results in a similar inequality with a redefined test function $\lambda(A,a)$.
    We define the vector $\boldsymbol{\lambda}$ as the set of elements corresponding to the redefined test function $\lambda(A,a)$.
    Consequently, the inequality (\ref{Eq:BellLikeInequalities}) can be rewritten as
	\begin{align}
		\label{Eq:VectorBLI}
		\boldsymbol{\mathcal{P}} \cdot \boldsymbol{\lambda} \leq \sup_{\alpha \in \mathbb{C}} \boldsymbol{\Pi}(\alpha) \cdot \boldsymbol{\lambda},
	\end{align}
	where $\boldsymbol{\Pi}(\alpha)$ represents the $Q$ symbols of the POVM, i.e., the probability distributions for the coherent states.

    Let us analyze inequality~(\ref{Eq:VectorBLI}) in the context of convex geometry; cf., e.g., Refs.~\cite{Hug_book,Rockafellar_book}.
    In the space of vectors $\boldsymbol{\mathcal{P}}$, the probability distributions for the coherent states (the $Q$ symbols of the POVM) form a manifold
        \begin{align}\label{Eq:Manifold}
           \mathcal{C} = \left\{ \boldsymbol{\Pi}(\alpha) \middle| \alpha \in \mathbb{C}_{\infty} \right\},
        \end{align}
    described by the parametric equation $\boldsymbol{\mathcal{P}}=\boldsymbol{\Pi}(\alpha)$, where $\mathbb{C}_{\infty}$ is the extended complex plane\footnote{The set $\mathcal{C}$ can be a manifold with boundary. For brevity, however, we refer to it as a manifold.}.
    In the most general case, the manifold $\mathcal{C}$ is two-dimensional, since it is parametrized by two real parameters related to the complex number $\alpha$.
	However, for phase-insensitive measurements, such as photocounting, $\boldsymbol{\Pi}(\alpha)$ depends only on $|\alpha|$ and not on $\arg\alpha$.
	This makes the corresponding manifold one-dimensional, i.e., it is a curve.
    The classical probability distributions (vectors) $\boldsymbol{\mathcal{P}}$ belong to the convex hull of the manifold $\mathcal{C}$,
        \begin{align}\label{Eq:Hull}
            \mathcal{H}=\Conv\mathcal{C}.
        \end{align}
    Nonclassicality of $\boldsymbol{\mathcal{P}}$ implies that $\boldsymbol{\mathcal{P}}{\notin}\mathcal{H}$.

\begin{figure}[t!]
    	\centering
    	\includegraphics[width=\linewidth]{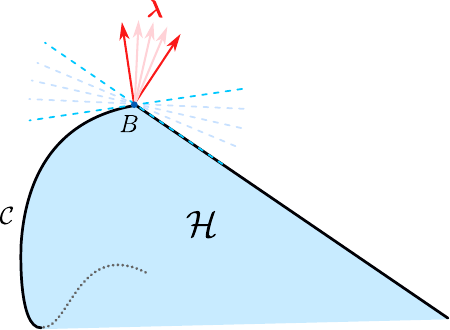}
    	\caption{\label{Fig:ToyExample} A toy example demonstrating the manifold $\mathcal{C}$ (solid and dotted lines) and its convex hull $\mathcal{H}$ (shaded area).
        At the breaking point $B\in\mathcal{C}$ there exist multiple supporting hyperplanes (dashed lines) corresponding to different vectors $\boldsymbol{\lambda}$.
        The points from the dotted part of $\mathcal{C}$ are not associated with any supporting hyperplane.}
        \end{figure}

    The right-hand side of Eq.~(\ref{Eq:VectorBLI}) can be interpreted as the support function \cite{Hug_book} of $\mathcal{H}$, see Refs.~\cite{Filip2013,Innocenti2022,Innocenti2023}.
    This function represents the distance (in the units of length of $\boldsymbol{\lambda}$) from the origin to the supporting hyperplane of $\mathcal{H}$ normal to the vector $\boldsymbol{\lambda}$.
    Each supporting hyperplane of $\mathcal{H}$ should include at least one point of the manifold $\mathcal{C}$ corresponding to a value of the parameter $\alpha$; see Appendix~\ref{App:GenStHP}.
    Hence, each vector $\boldsymbol{\lambda}$ can be associated with at least one value of the parameter $\alpha$.
    However, on the manifold $\mathcal{C}$ there can exist such points which (i) have more than one supporting hyperplanes of $\mathcal{H}$ associated with each of them or (ii) do not belong to any supporting hyperplane of $\mathcal{H}$; see Fig.~\ref{Fig:ToyExample} for a toy example.
    This leads to the necessity of optimizing the set of vectors $\boldsymbol{\lambda}$ in the context of Definition~\ref{DefTight} and the procedure of their construction.

	The standard methods of convex analysis (see, e.g., Refs.~\cite{Rockafellar_book, Hug_book}) yield a statement that serves as a practical tool for constructing tight inequalities.
	Prior to introducing this statement, let us recall the concept of closed convex cones and their extreme rays.
    A closed set $L$ of vectors is a closed convex cone if for any $\boldsymbol{\lambda}_1, \boldsymbol{\lambda}_2\in L$ the conic combination $C_1\boldsymbol{\lambda}_1+C_2\boldsymbol{\lambda}_2\in L$, where $C_{1,2}\geq0$.
    The extreme rays of $L$, denoted as $\Ext\left[ L\right]$, comprise a continuous, discrete, or hybrid set of vectors $\boldsymbol{\lambda}_i$ such that any $\boldsymbol{\lambda} \in L$ can be expressed as a conic combination of these vectors, i.e., $\boldsymbol{\lambda} = \sum_i C_i \boldsymbol{\lambda}_i$, where $C_i \geq 0$, and sum is replaced by integral for continuous $i$.
    Additionally, no vector $\lambda_k\in\Ext \left[ L\right]$ can be represented as a conic combination of other vectors in this set.
    The statement used for constructing tight inequalities can now be formulated in the following form:

		\begin{statement}
			\label{St:TightCondition0}
            Let us consider the closed convex cones, cf. Appendix~\ref{App:GenStCC}, defined by
                \begin{align}\label{Eq:LambdaSetFormula0}
                    L(\alpha) = \left\{ \boldsymbol{\lambda} \middle| \argmax_{\alpha^\prime \in \mathbb{C}_{\infty}}  \boldsymbol{\Pi}(\alpha^\prime) \cdot\boldsymbol{\lambda} = \alpha \right\}.
                \end{align}
			The set $\Lambda$, producing tight inequalities, cf. Definition~\ref{DefTight}, can be obtained as
			\begin{align}
			\label{Eq:LambdaSetFormula}
			\Lambda =  \bigcup_{\alpha \in \mathbb{C}_{\infty}} \Ext \left[L(\alpha)\right].
			\end{align}
		\end{statement}

    The proof of this Statement is given in Appendix~\ref{App:GenStSt1}.
    Its essence can be explained as follows.
    Equation~(\ref{Eq:LambdaSetFormula0}) assigns, for each value of $\alpha$, the set $L(\alpha)$ of vectors $\boldsymbol{\lambda}$ for which the expression $\boldsymbol{\Pi}(\alpha^{\prime})\cdot\boldsymbol{\lambda}$ reaches the global maximum as a function of $\alpha^{\prime}$.
    There can be such values of $\alpha$ for which global maximum is not reached for any vector $\boldsymbol{\lambda}$.
    In such cases, $L(\alpha)=\varnothing$.
    In another situation, a given value of $\alpha$ can be assigned to multiple vectors $\boldsymbol{\lambda}$ for which the global maximum is reached.
    The extreme ray optimizes the set $L(\alpha)$.
    Finally, Eq.~(\ref{Eq:LambdaSetFormula}) unites the optimized sets of $\boldsymbol{\lambda}$ for different values of $\alpha$.

	\begin{figure}[t!]
	\includegraphics[width=\linewidth]{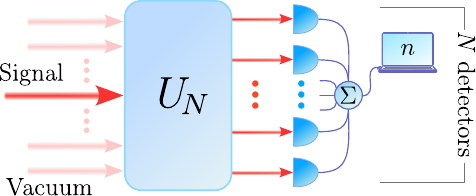}
	\caption{\label{Fig:Array} A principal scheme of a click detector based on the spatial splitting of a light mode, represented by a unitary $U(N)$, into $N$ equally balanced modes, each of them is analyzed by an on-off detector.
		The rest $N-1$ modes at the input of the interferometer are in the vacuum state.
		The outcome of the detection system is given by the number $n=0\ldots N$ of all triggered detectors.}
	\end{figure}

\section{Photocounting measurements}
\label{Sec:PC}
	First, we consider photocounting, which is a typical example of quantum measurements.
	In this case, the measurement outcomes correspond to the integer values of $A=n$.
	Depending on the detection scheme used, the number $n$ can be interpreted as the number of detected photons in the case of PNR detectors, or as the number of clicks in the case of detection schemes with imperfect photon-number resolution.
	Since this measurement scheme does not have device settings---i.e., the parameter $a$ has only a single value, $N_{\mathrm{a}}=1$ in Eq.~(\ref{Eq:Dimensionality})---we omit them here.

	For the PNR detectors, the $Q$ symbols of the POVM are given by (see Refs.~\cite{mandel_book, kelley64})
	\begin{align}\label{Eq:IdealDetectPOVM}
		\Pi(n|\alpha)=\frac{|\alpha|^{2n}}{n!}\exp\left(-|\alpha|^2\right).
	\end{align}
	Let us restrict our analysis to $n=0\ldots N-1$, where $N$ is an integer.
	We also consider that the outcome with $n{=}N$ corresponds to all events with $N$ and more detected photons, which implies
	\begin{align}
		\Pi(N|\alpha)=1-\sum\limits_{n=0}^{N-1}\Pi(n|\alpha).
	\end{align}
	This POVM element depends on the rest of the elements.
	The measurement device has $N_{\mathrm{A}}=N+1$ outcomes.
    According to Eq.~(\ref{Eq:Dimensionality}), $\dim\boldsymbol{\mathcal{P}}=N$.

    A finite number of outcomes, $N_{\mathrm{A}}$, is usually inherent to realistic photocounting measurements; see, e.g., Refs.~\cite{sperling12a,Uzunova2022,Semenov2024,Stolyarov2023}.
    In this paper, we focus on a particular example of such detection schemes, so-called click detectors.
    The corresponding layouts are based on spatial \cite{paul1996,castelletto2007,schettini2007,blanchet08} or temporal \cite{achilles03,fitch03,rehacek03} splitting of the light beam into several modes, followed by separate detection of each mode by on-off detectors.
	A typical scheme is shown in Fig.~\ref{Fig:Array}.
	The measurement outcome $n$ in this case corresponds to the total number of triggered detectors (clicks).
	The corresponding $Q$ symbols of the POVM read 
	\begin{align}
		\label{Eq:ArrayDetectorsPOVM}
		\Pi(n|\alpha) = \binom{N}{n} e^{-(N - n)|\alpha|^2/N} \left( 1 - e^{-|\alpha|^2 / N} \right)^n,
	\end{align}
	where $N$ represents the number of on-off detectors and $n=0\ldots N$; see Ref.~\cite{sperling12a}.
	Similar to the case of the PNR detectors, the element $\Pi(N|\alpha)$ can be expressed in terms of other elements---i.e., $N_{\mathrm{A}}=N+1$---and thus, as it follows from Eq.~(\ref{Eq:Dimensionality}), $\dim\boldsymbol{\mathcal{P}}=N$.
    In all cases, we consider the detection efficiency to be unity and attribute the detection losses to the quantum states.

	Since all photocounting measurements correspond to phase-insensitive measurements, the $Q$ symbols of the POVM, denoted as $\boldsymbol{\Pi}(\alpha)$, do not depend on the phase of $\alpha$.
	To conveniently parametrize them, we introduce the parameter $t=e^{-|\alpha|^2/N}\in [0,1]$.
	Evidently, the manifold $\mathcal{C}$, cf. Eq.~(\ref{Eq:Manifold}), in this case forms an open curve in an $N$-dimensional space, with the beginning and endpoints corresponding to $t=0$ and $t=1$, respectively.
	For the sake of brevity, we also adopt the symbol $\boldsymbol{\Pi}(t)$ to represent the dependence of the $Q$ symbols of the POVM in photocounting measurements on the parameter $t$.

\subsection{Photocounting with $N{=}2$}
\label{Sec:PC2}

	The case of $N{=}1$, in both cases of the POVMs (\ref{Eq:IdealDetectPOVM}) and (\ref{Eq:ArrayDetectorsPOVM}), corresponds to a single on-off detector.
	This scenario is not capable for testing nonclassicality \cite{sperling12c,Semenov2021}.
	For this reason, we start our consideration with the simplest nontrivial case of $N{=}2$, corresponding to two-dimensional vectors $\boldsymbol{\mathcal{P}}$; cf. Eq.~(\ref{Eq:Dimensionality}) with $N_{\mathrm{A}}=3$ and $N_{\mathrm{a}}=1$.
	Evidently, the set $\mathcal{C}$ of the $Q$ symbols of the POVM is represented in this case by the open curve in a two-dimensional space, given by the parametric equations
	\begin{align}
		\mathcal{P}(0){=}\Pi(0|t),\label{Eq:Pram1}\\
		\mathcal{P}(1){=}\Pi(1|t).\label{Eq:Pram2}
	\end{align}
	Inverting $\Pi(0|t)$ as a function of $t$ and substituting the result into Eq.~(\ref{Eq:Pram2}), we arrive at the explicit equation for this curve,
	\begin{align}\label{Eq:Curve}
		\mathcal{P}(1){=}\Pi_1(\mathcal{P}(0)),
	\end{align}
	where the function $\Pi_1(\mathcal{P}(0))$ depends on the detection scheme used.
	We restrict our consideration to the detectors defined by the POVMs (\ref{Eq:IdealDetectPOVM}) and (\ref{Eq:ArrayDetectorsPOVM}).
	However, this analysis is also applicable to a broader class of detectors for which (i) $\Pi(0|t{=}0){=}\Pi(1|t{=}0){=}0$, i.e., the detector is saturated at $|\alpha|^2\rightarrow+\infty$; (ii) $\Pi(0|t{=}1){=}1$, $\Pi(1|t{=}1){=}0$, i.e., no dark counts are present; (iii) the first derivatives $\dot{\Pi}(0|t)$ and $\dot{\Pi}(1|t)$ exist for $t\in[0,1]$; (iv) the second derivative of $\Pi_1(x)$ obeys $\Pi_1^{\prime\prime}(x)\leq0$.

    \begin{figure}[t!]
	\includegraphics[width=\linewidth]{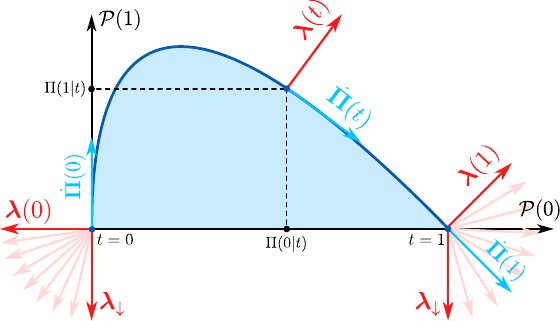}
	\caption{\label{Fig:2D} The solid line corresponds to the function $\boldsymbol{\mathcal{P}}=\boldsymbol{\Pi}(t)$ (i.e., the manifold $\mathcal{C}$) for the click detectors with $N=2$.
		Its convex hull $\mathcal{H}$ is marked by the shaded area.
		The vector $\boldsymbol{\lambda}(t)$ for $t\in[0,1]$ is normal to the vector $\dot{\boldsymbol{\Pi}}(t)$.
		At the endpoints $t=0$ and $t=1$, the maximum of $\boldsymbol{\Pi}(t)\cdot\boldsymbol{\lambda}$ is also reached for $\boldsymbol{\lambda}=\boldsymbol{\lambda}_{\downarrow}$ and their conic combinations (pale-colored vectors) with the vectors $\boldsymbol{\lambda}(0)$ and $\boldsymbol{\lambda}(1)$ for $t=0$ and $t=1$, respectively.}
	\end{figure}

	The direct application of Statement~\ref{St:TightCondition0} in this case implies finding $\boldsymbol{\lambda}(t)$ from the equation 
    \begin{align}\label{Eq:Maxt1}
        \dot{\boldsymbol{\Pi}}(t)\cdot\boldsymbol{\lambda}(t)=0
    \end{align}
    for each $t{\in}]0,1[$.
    For this open interval, this equation follows from the condition 
    \begin{align}\label{Eq:Maxt0}
        \argmax_{t^\prime \in [0,1]}\boldsymbol{\Pi}(t^\prime) \cdot\boldsymbol{\lambda} = t
    \end{align}
    from Statement~\ref{St:TightCondition0}.
    Since the solutions to this equation are collinear vectors $\boldsymbol{\lambda}(t)$, the corresponding extreme ray consists of a single vector,
        \begin{align}\label{Eq:Vec2D}
			\boldsymbol{\lambda}(t)=\begin{pmatrix}-\dot{\Pi}(1|t) & \dot{\Pi}(0|t)\end{pmatrix},
		\end{align}
     normal to the supporting hyperplane (line) at the point $t$.
    
    The situation is different at the endpoints $t=t_{\mathrm{end}}$, where $t_{\mathrm{end}}$ takes values 0 and 1; cf. Fig.~\ref{Fig:2D}.
     For these points, condition (\ref{Eq:Maxt0}) is satisfied for any conic combination of vectors $\boldsymbol{\lambda}=\boldsymbol{\lambda}(t_{\mathrm{end}})$ and $\boldsymbol{\lambda}=\boldsymbol{\lambda}_{\downarrow}$. 
     Here $\boldsymbol{\lambda}_{\downarrow}=\begin{pmatrix}0 & -1\end{pmatrix}$.
     Indeed, for $\boldsymbol{\lambda}=\boldsymbol{\lambda}(t_{\mathrm{end}})$ this condition is fulfilled in virtue of Eq.~(\ref{Eq:Maxt1}).
     At the same time, the maximum value of $\boldsymbol{\lambda}_{\downarrow} \cdot \boldsymbol{\Pi}(t)$ equals $0$ and is reached at $t = t_{\mathrm{end}}$, since $\boldsymbol{\lambda}_{\downarrow} \cdot \boldsymbol{\Pi}(t) = -\Pi(0|t) < 0$ for $t\in]0,1[$.
     Evidently, the conic combinations of vectors $\boldsymbol{\lambda}(t_{\mathrm{end}})$ and $\boldsymbol{\lambda}_{\downarrow}$ reach maximum at $t=t_{\mathrm{end}}$ as well.
     This yields $\Ext  \left[L(t{=}t_{\mathrm{end}})\right]=\{\boldsymbol{\lambda}(t_{\mathrm{end}}),\boldsymbol{\lambda}_{\downarrow}\}$.
     Hence, the vector $\boldsymbol{\lambda}_{\downarrow}$ together with vectors (\ref{Eq:Vec2D}) for $t=[0,1]$ determine the set of tight inequalities according to Eq.~(\ref{Eq:LambdaSetFormula}) from Statement~\ref{St:TightCondition0}. 
    However, the vector $\boldsymbol{\lambda}_{\downarrow}$ results in a trivial inequality for non-negativity of $\mathcal{P}(1)$.
   
	Therefore, the set of tight inequalities for the considered case is given by
	\begin{align}
		\boldsymbol{\mathcal{P}}\cdot\boldsymbol{\lambda}(t)\leq\boldsymbol{\Pi}(t)\cdot\boldsymbol{\lambda}(t).
	\end{align}
	This set of linear inequalities can be combined into a single nonlinear inequality as
	\begin{align}\label{Eq:IneqDetectors}
		\mathcal{P}(1)\leq\Pi_1(\mathcal{P}(0)).
	\end{align}
	In the case of PNR detectors, cf. Eq.~(\ref{Eq:IdealDetectPOVM}), this leads to inequality $\mathcal{P}(1){\leq}{-}\mathcal{P}(0)\ln\mathcal{P}(0)$ derived with a different technique in Ref.~\cite{Filip2013}.
	In the case of click detectors, cf.~Eq.~(\ref{Eq:ArrayDetectorsPOVM}), inequality (\ref{Eq:IneqDetectors}) reduces to the form
	\begin{align}\label{Eq:NonlinearPD}
		\left[2\mathcal{P}(0) + \mathcal{P}(1)\right]^2 - 4\mathcal{P}(0) \leq 0,
	\end{align}
	see Appendix~\ref{App:N2} for details.
	Violation of this inequality is a necessary and sufficient condition for nonclassicality of the probability distribution $\boldsymbol{\mathcal{P}}$ and a sufficient condition for nonclassicality of the corresponding quantum state.

	We demonstrate the applicability of this method with the statistics of the attenuated Fock state $\ket{n}$.
	In this scenario inequality (\ref{Eq:NonlinearPD}) takes the form $(1-\eta/2)^{2n}\leq(1-\eta)^n$, where $\eta\in]0,1]$ is the efficiency.
	This inequality is always violated, although for large $n$ the detector is saturated and the violation becomes small.
    Hence $\boldsymbol{\mathcal{P}}{\notin}\mathcal{H}$, which implies that these photocounting statistics are nonclassical.

	\begin{figure}[t!]
	\includegraphics[width=\linewidth]{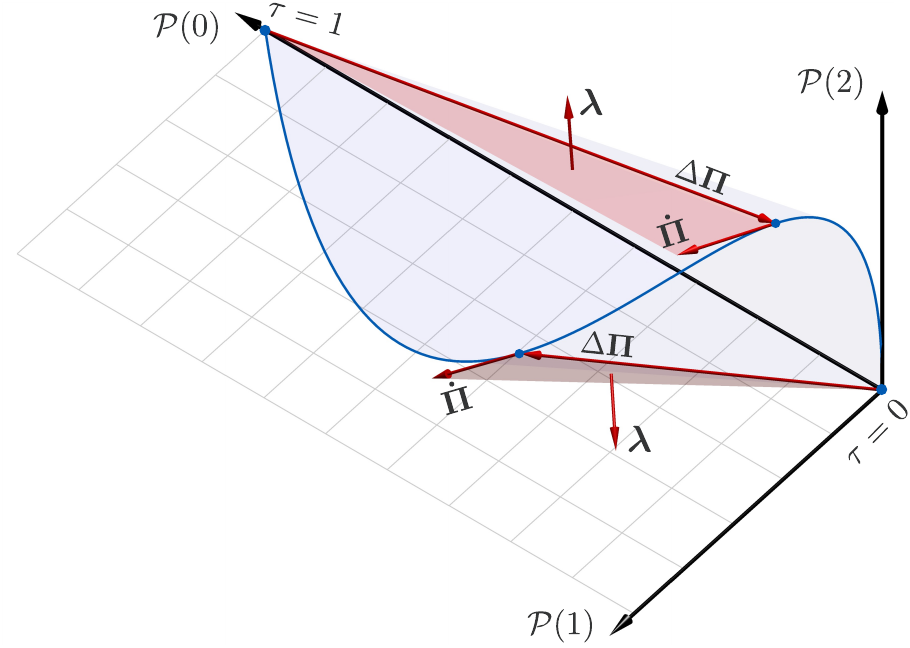}
	\caption{\label{Fig:3D} The curve $\mathcal{C}$ and its convex hull $\mathcal{H}$ (shaded volume) for the photocounting measurement with click detectors in the case of $N=3$ is shown.
		The vectors $\dot{\boldsymbol{\Pi}}(t_1)$ and $\Delta \boldsymbol{\Pi}(t_1,\tau)$ for two different values of $t_1$ and for $\tau=0$ and $\tau=1$ form supporting planes of the convex hull of the curve.
		The corresponding vectors $\boldsymbol{\lambda}$ are normal to these planes.}
	\end{figure}

\subsection{Photocounting with $N=3$}
\label{Sec:PC3}

    The larger number of possible measurement outcomes enables us to verify nonclassicality of photocounting statistics for a broader range of quantum states.
   	However, if the dimensionality of the vector $\boldsymbol{\mathcal{P}}$ exceeds two, the task of constructing the convex hull $\mathcal{H}$ of the curve $\mathcal{C}$ defined by $\boldsymbol{\mathcal{P}}=\boldsymbol{\Pi}(t)$ becomes more difficult compared with the case of $N=2$; see Ref.~\cite{Pont2023} for a recent mathematical result in the field.
    In this paper, we use a heuristic method to construct a set of supporting hyperplanes of $\mathcal{H}$ and then prove that the associated vectors $\boldsymbol{\lambda}$ correspond to tight inequalities in the context of Definition~\ref{DefTight}.
    We first demonstrate the approach for $N=3$ and then generalize it to arbitrary odd $N$.

    In the considered scenario, there are three linearly independent components of the vector $\boldsymbol{\mathcal{P}}$: $\mathcal{P}(0)$, $\mathcal{P}(1)$, and $\mathcal{P}(2)$.
    Thus, $\mathcal{C}$ is a curve in $\mathbb{R}^3$, see Fig.~\ref{Fig:3D}.
    First, we use the fact that any supporting plane of $\mathcal{H}$ containing the point $\boldsymbol{\Pi}(t_1)$, where $t_1\in[0,1]$, is parallel to the vector $\dot{\boldsymbol{\Pi}}(t_1)$ tangent to the curve $\mathcal{C}$.
    Next, we assume that the required supporting plane contains one of the endpoints of this curve, $\boldsymbol{\Pi}(\tau)$, where $\tau = 0$ or $\tau = 1$, and $t_1\neq\tau$.
    Therefore, for each point $t=t_1$, we assign two planes related to different values of $\tau$. The corresponding normal vectors are defined via the cross product as
        \begin{align}\label{Eq:TightLambda3D}
            \boldsymbol{\lambda}(t_1;\tau) =\pm \Delta \boldsymbol{\Pi}(t_1,\tau)\times \dot{\boldsymbol{\Pi}}(t_1),
        \end{align}
    where $\Delta \boldsymbol{\Pi}(t_1,\tau)=\boldsymbol{\Pi}(t_1) - \boldsymbol{\Pi}(\tau)$.
    In the next section we prove that the constructed planes are supporting planes of $\mathcal{H}$ and there exists such a sign in Eq.~(\ref{Eq:TightLambda3D}) that the vectors $\boldsymbol{\lambda}(t_1;\tau)$ correspond to tight inequalities.
    This statement, however, is proved for a restricted class of detection schemes, including the PNR and click detectors, cf. Eqs.~(\ref{Eq:IdealDetectPOVM}) and (\ref{Eq:ArrayDetectorsPOVM}).

    In the considered case of $N=3$, the set of tight inequalities can be combined into two nonlinear inequalities.
    For the PNR detectors, cf.~Eq.~(\ref{Eq:IdealDetectPOVM}), these inequalities are given by
        \begin{align}
            &\mathcal{P}^2(1)\leq 2\mathcal{P}(0)\mathcal{P}(1),\label{Eq:PNR3_1}\\
            &\mathcal{P}(0)+\frac{\mathcal{P}^2(1)}{2\mathcal{P}(2)}\left[\exp\left(\frac{2\mathcal{P}(2)}{\mathcal{P}(1)}\right)-1\right]\leq 1.\label{Eq:PNR3_2}
        \end{align}
    Inequalities (\ref{Eq:PNR3_1}) and (\ref{Eq:PNR3_2}) are obtained from the tight inequalities with $\tau=0$ and $\tau=1$, respectively.
    The same inequalities have been derived in Ref.~\cite{Innocenti2022} with a different technique.
    Moreover, inequality (\ref{Eq:PNR3_1}) is, in fact, a Klyshko inequality \cite{klyshko1996}. 
    For the click detectors, cf.~Eq.~(\ref{Eq:ArrayDetectorsPOVM}), the corresponding inequalities read
        \begin{align}
            &\mathcal{P}^2(1)\leq 3\mathcal{P}(0)\mathcal{P}(2),\label{Eq:Array3_1}\\
            & 3 \mathcal{P}^2(1)+\mathcal{P}^2(2)+3\mathcal{P}(1)\left[\mathcal{P}(0)+\mathcal{P}(2)-1\right]\leq 0.\label{Eq:Array3_2}
        \end{align}
    Similar to the case of the PNR detectors, inequalities (\ref{Eq:Array3_1}) and (\ref{Eq:Array3_2}) are derived from the tight inequalities with $\tau=0$ and $\tau=1$, respectively.
    For the derivation of these inequalities, see Appendix~\ref{App:N3}.

	\begin{figure}[t!]
	\includegraphics{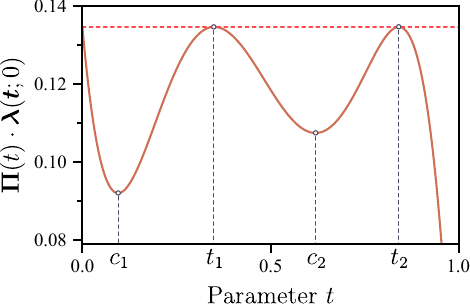}
	\caption{\label{Fig:CrPoints} An example of $\boldsymbol{\Pi}(t) \cdot \boldsymbol{\lambda}(\boldsymbol{t};\tau=0)$ as a function of $t$ for click detectors with $N=5$ (i.e., $m=2$) is shown. The sets of points $0$, $t_1$, $t_2$ and $c_1$, $c_2$ correspond to global maxima and local minima, respectively.}
	\end{figure}

\subsection{Photocounting with $N=2m+1$, $m\in\mathbb{N}$}
\label{Sec:Odd}

	Let us generalize our analysis to $N=2m+1$, where $m\in\mathbb{N}$.
    According to Eq.~(\ref{Eq:Dimensionality}), $\dim\boldsymbol{\mathcal{P}}=N$ and thus the dimensionality of the supporting hyperplanes of $\mathcal{H}$ is equal to $2m$.
  	Similar to the case of $N=3$ we heuristically construct hyperplanes and then prove that they are supporting hyperplanes of $\mathcal{H}$.

	\begin{figure}[t!]
	\includegraphics{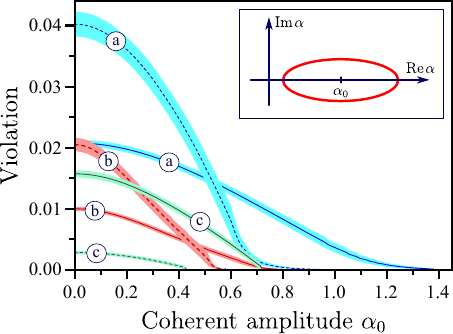}
	\caption{\label{Fig:OddN} The maximum violation of the tight inequalities (difference of their left- and right-hand sides optimized by parameters $\boldsymbol{t}$ and $\tau$) with the test function given by Eq.~(\ref{Eq:OddPDTightLambda}) for photocounts of phase-squeezed coherent states vs the coherent amplitude $\alpha_0$ is shown; the Wigner function of such a state is mimicked by its contour line in the inset.
		The confidence intervals correspond to the standard deviations for estimations of inequality left-side with $10^5$ sample events.
		The solid and dashed lines are related to the PNR and click detectors, respectively.
		The overall losses correspond to efficiency $\eta=0.7$.
		The plots describe the cases: (a) $N=5$, $r=0.57$; (b) $N=5$, $r=0.34$; (c) $N=3$, $r=0.57$. }
	\end{figure}

    First, we require that each hyperplane we construct contains $m$ points $\boldsymbol{\Pi}(t_i)$, where $i = 1\ldots m$, such that $0 \leq t_1 < t_2 < \ldots < t_m \leq 1$.
    For this hyperplane to be a supporting hyperplane of $\mathcal{H}$, it is necessary that it be parallel to the $m$ vectors $\dot{\boldsymbol{\Pi}}(t_i)$ tangent to the curve $\mathcal{C}$.
    Next, we assume that it also contains one of the endpoints of this curve, $\boldsymbol{\Pi}(\tau)$, where $\tau = 0$ or $\tau = 1$, and $t_i \neq \tau$ for all $i$.
    Similar to the case of $N=3$, we define two hyperplanes (for different values of $\tau$), containing the points $\boldsymbol{t}=\{t_1,\ldots t_m\}$.
    The corresponding normal vectors are given by
		\begin{align}\label{Eq:OddPDTightLambda}
		\boldsymbol{\lambda}(\boldsymbol{t};\tau) = \pm \left( \star \bigwedge_{i = 1}^m \Delta \boldsymbol{\Pi}(t_i,\tau) \wedge \dot{\boldsymbol{\Pi}}(t_i)\right).
		\end{align}
	Here $\Delta\boldsymbol{\Pi}(t_i,\tau) = \boldsymbol{\Pi}(t_i) - \boldsymbol{\Pi}(\tau)$ and $\star$ is the Hodge star, which is a generalization of the cross product, see Ref.~\cite{Frankel_book} and Appendix~\ref{App:Hodge}.
	These vectors fulfill the conditions,
       \begin{align}
            &\Delta\boldsymbol{\Pi}(t_i,\tau)\cdot\boldsymbol{\lambda}(\boldsymbol{t};\tau) = 0,\label{Eq:Cond1}\\
            &\dot{\boldsymbol{\Pi}}(t_i) \cdot \boldsymbol{\lambda}(\boldsymbol{t};\tau) = 0,\label{Eq:Cond2}
        \end{align}
    for $i=1\ldots m$.
    These equations can be obtained by directly substituting the vector $\boldsymbol{\lambda}$ from Eq.~(\ref{Eq:OddPDTightLambda}) and applying Eq.~(\ref{Eq:Hodge}).
    The equations indicate that the vector $\boldsymbol{\lambda}(\boldsymbol{t};\tau)$ is orthogonal to the vectors $\dot{\boldsymbol{\Pi}}(t_i)$ and $\Delta\boldsymbol{\Pi}(t_i,\tau)$.
    Also Eq. ~(\ref{Eq:Cond1}), rewritten as $\boldsymbol{\Pi}(t_i)\cdot\boldsymbol{\lambda}(\boldsymbol{t};\tau) =\boldsymbol{\Pi}(\tau)\cdot\boldsymbol{\lambda}(\boldsymbol{t}; \tau)$, implies that the values of $\boldsymbol{\Pi}(t)\cdot\boldsymbol{\lambda}(\boldsymbol{t};\tau)$ are equal for all $t=t_i$ and $t=\tau$.

    In general, the vectors $\boldsymbol{\lambda}(\boldsymbol{t};\tau)$ given by Eq.~(\ref{Eq:OddPDTightLambda}) may not lead to tight inequalities for arbitrary detection schemes defined by $\boldsymbol{\Pi}(t)$.
    However, we prove that for a class of detectors including the PNR and click detectors characterized by (\ref{Eq:IdealDetectPOVM}) and (\ref{Eq:ArrayDetectorsPOVM}), respectively, these vectors define a subset of tight inequalities.
    Equations~(\ref{Eq:Cond1}) and (\ref{Eq:Cond2}) represent a set of $2m$ linear constraints, which, if not degenerate, define $\boldsymbol{\lambda}(\boldsymbol{t};\tau)$ given by Eq.~(\ref{Eq:OddPDTightLambda}) up to a constant factor.
    These equations are also necessary but not sufficient conditions for $\boldsymbol{\lambda}(\boldsymbol{t};\tau)\in \Lambda$.
    Indeed, Eq.~(\ref{Eq:Cond2}) does not imply that the function $\boldsymbol{\Pi}(t_i) \cdot \boldsymbol{\lambda}(\boldsymbol{t};\tau)$ does not reach a maximum at the points $t_i$ and even if it did, there would still be no guarantee that it is global.
    For a wide class of detectors, however, the global maximum of the above function is reached at the points $t_i$ and $\tau$; cf.~Fig.~\ref{Fig:CrPoints}.
    This conclusion is the essence of the following statement.

    \begin{statement}\label{St:PhotoCount1}
	Apart from $t_1,\ldots t_m$, let $\boldsymbol{\Pi}(t) \cdot \boldsymbol{\lambda}(\boldsymbol{t}, \tau)$ have $m$ other critical points at $t \in (0, 1)$.
	Furthermore, $ \ddot{\boldsymbol{\Pi}}(t) \cdot \boldsymbol{\lambda}(\boldsymbol{t}; \tau)$ has $2m - 1$ zeros at $t \in (0, 1)$.
	Then one can choose such a sign of $\boldsymbol{\lambda}(\boldsymbol{t};\tau)$ in Eq.~(\ref{Eq:OddPDTightLambda}), that the global maximum of  $\boldsymbol{\Pi}(t) \cdot \boldsymbol{\lambda}(\boldsymbol{t};\tau)$ is attained for $t=t_i$ and $t=\tau$.
    \end{statement}

    The proof of this statement can be found in Appendix~\ref{App:Prof-St2}.
    In Appendix~\ref{App:CondDet}, it is shown that the PNR and click detectors defined by Eqs.~(\ref{Eq:IdealDetectPOVM}) and (\ref{Eq:ArrayDetectorsPOVM}), respectively, satisfy the conditions of this statement.
    Therefore, for each point $t$ we have found a set of test functions $\boldsymbol{\lambda}$, such that the $\boldsymbol{\Pi}(t) \cdot \boldsymbol{\lambda}$ reaches the global maximum.
    Hence, the hyperplanes normal to the vector $\boldsymbol{\lambda}(\boldsymbol{t};\tau)$ given by Eq.~(\ref{Eq:OddPDTightLambda}) represent supporting hyperplanes of $\mathcal{H}$.

  	Next, we show that every $\boldsymbol{\lambda}(\boldsymbol{t};\tau)$ defined by Eq.~(\ref{Eq:OddPDTightLambda}) belongs to $\Ext\left[ L(t_k)\right]$ for every $t_k$.
	A similar statement can be proved for $\Ext\left[ L(\tau)\right]$.
	Suppose $\boldsymbol{\lambda}(\boldsymbol{t};\tau)\notin\Ext\left[ L(t_k)\right]$.
	Then, there exist such $\boldsymbol{\lambda}_1, \boldsymbol{\lambda}_2\in L(t_k)$ that $\boldsymbol{\lambda}(\boldsymbol{t};\tau)$ is their conic combination; cf. Ref.~\cite{Rockafellar_book} and Appendix~\ref{App:ExtrRays}.
	In this case it can be shown that they fulfill the above conditions (\ref{Eq:Cond1}) and (\ref{Eq:Cond2}) for $\boldsymbol{\lambda}(\boldsymbol{t};\tau)$.
	These conditions represent a nondegenerate system of $N$ linear homogeneous equations for $N$ components of $\boldsymbol{\lambda}$.
	The solutions to this system are given by the collinear vectors.
	Thus, $\boldsymbol{\lambda}(\boldsymbol{t};\tau)$ cannot be expanded with two other noncollinear vectors.
	This means that $\boldsymbol{\lambda}(\boldsymbol{t};\tau)\in\Ext\left[ L(t_k)\right]$; cf. Ref.~\cite{Rockafellar_book} and Appendix~\ref{App:ExtrRays}.
	According to Statement~\ref{St:TightCondition0}, this implies that the vectors $\boldsymbol{\lambda}(\boldsymbol{t};\tau)$, given by Eq.~(\ref{Eq:OddPDTightLambda}), belong to a set of tight inequalities.
	However, we do not prove here that they form the complete set of these inequalities.

	We demonstrate the proposed method with phase-squeezed coherent states $\ket{\alpha_0,r}=\hat{D}(\alpha_0)\hat{S}(r)\ket{0}$. Here, $\hat{D}(\alpha_0)$ is the displacement operator with $\alpha_0\in\mathbb{R}$, $\hat{S}(r)$ is the squeeze operator with $r>0$, and $\ket{0}$ is the vacuum state; cf. the inset in Fig.~\ref{Fig:OddN} for a contour line of the corresponding Wigner function.
    Such states exhibit super-Poissonian statistics of photocounts.
    Phase-insensitive measurements, like photocounting, and measurements of their amplitude quadrature are typically considered unsuitable for testing their nonclassicality \cite{Schnabel2017}.

	The results in Fig.~\ref{Fig:OddN} challenge the prevailing intuition about the classical character of photocounting statistics for phase-squeezed coherent states.
    Contrary to this intuition, the statistics are found to be nonclassical, as evinced by violations of tight inequalities with $N=3$ and $N=5$ for both PNR and click detectors.
    While the Klyshko criterion can also demonstrate this nonclassicality, it is limited to PNR detectors.
    In contrast, our method can be employed with experimentally accessible click detectors.

\section{Unbalanced homodyne detection}
\label{Sec:UHD}
\begin{figure}[t!]
            \includegraphics{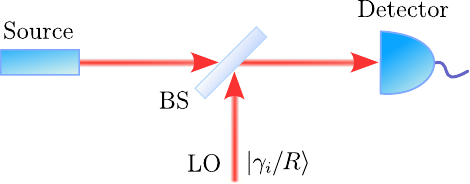}
	 	\caption{\label{Fig:UHD_Scheme}  The scheme of unbalanced homodyne detection is shown.
        The signal field is superposed with the local-oscillator (LO) field in the coherent state $\ket{\gamma_i/R}$ at a beam splitter (BS) with the transmission coefficient $T\approx 1$, where $R$ is the reflection coefficient of the BS.
        The resulting light field, displaced in phase space, is analyzed by an on-off detector.
        The amplitude $\gamma_i$ of the LO plays the role of the device setting.}
	   \end{figure}

	As mentioned, the binary measurement outcomes obtained with on-off detectors ($N_{\mathrm{A}}=2$, i.e., $N=1$) cannot be used for testing nonclassicality.
	However, the situation changes when the signal field is superposed with the local-oscillator field, whose complex amplitude plays the role of the device setting.
	The corresponding measurement procedure of unbalanced homodyne detection \cite{wallentowitz96}, cf. Fig.~\ref{Fig:UHD_Scheme}, is described by the $Q$ symbols of the POVM given by
		\begin{align}\label{Eq:UHD_POVM}
		\Pi(0|\alpha;\gamma_i) = \exp\left(-|\alpha-\gamma_i|^2\right),
		\end{align}
	where $\gamma_i$ is the local-oscillator amplitude (the device setting).
	The dimensionality of the vector $\boldsymbol{\mathcal{P}}$, according to Eq.~(\ref{Eq:Dimensionality}) with $N_{\mathrm{A}}=2$, is equal to the number of settings, $N_{\mathrm{a}}$.
	We chose $N_{\mathrm{a}}=2$, which implies $\dim\boldsymbol{\mathcal{P}}=2$, and denote $d=|\gamma_2 - \gamma_1| / 2$.

	Similar to the examples above, all losses are assigned to quantum states.
	Another experimental imperfection, mode mismatch, according to its consideration in Ref.~\cite{banaszek2002}, leads to a multiplication of probabilities $\mathcal{P}(0|\gamma_i)$ and $\Pi(0|\alpha;\gamma_i)$ by a constant factor, depending on $\gamma_i$.
    It can be easily included in the consideration, cf.~Appendix~\ref{App:ModeMismatch}, and even omitted, if $|\gamma_1|=|\gamma_2|$.

\begin{figure}[t!]
				\centering
                \includegraphics{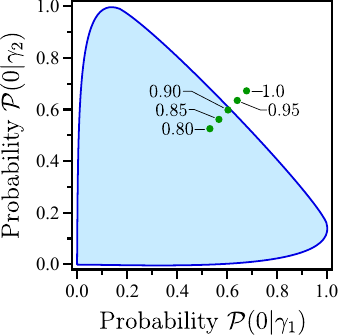}
				\caption{\label{Fig:UHD} The closed curve $\partial\mathcal{C}$ bounding the shaded two-dimensional region $\mathcal{H}$ of classical probability distributions is shown for unbalanced homodyne detection with two local-oscillator settings, $\gamma_{1,2}=\pm 1/\sqrt{2}$.
				The dots represent the probability distributions for the squeezed vacuum states with $r=0.34$.
                The numbers assigned to the dots are the values of the efficiency $\eta$.
				The dots outside $\mathcal{H}$ correspond to nonclassical probability distributions.
				}
			\end{figure}

	In the considered case, the manifold $\mathcal{C}$, cf. Eq.~(\ref{Eq:Manifold}), is given by a two-dimensional region, shown in Fig.~\ref{Fig:UHD}, bounded by the closed curve $\partial\mathcal{C}$ parametrically defined as
	\begin{align}
	\mathcal{P}(0|\gamma_1)=\Pi(0|t;\gamma_1)=\exp\left[-(t+d)^2\right],\label{Eq:UHD-Bound-1}\\
	\mathcal{P}(0|\gamma_2)=\Pi(0|t;\gamma_2)=\exp\left[-(t-d)^2\right]\label{Eq:UHD-Bound-2},
	\end{align}
	where $t\in\mathbb{R}$, see Appendix~\ref{App:Boundary}.
    The convex hull of any two-dimensional region is the convex hull of its boundary line.
    In our case, this means that $\Conv \mathcal{C}=\Conv\partial\mathcal{C}$.
    This statement follows from the expression,
        \begin{align}\label{Eq:BoundSup}
            \sup_{\alpha \in \mathbb{C}}\boldsymbol{\Pi}(\alpha)\cdot\boldsymbol{\lambda}=\sup_{t \in \mathbb{R}}\boldsymbol{\Pi}(t)\cdot\boldsymbol{\lambda},
        \end{align}
     see  Appendix~\ref{App:BounSup}.
     This implies that 
        \begin{align}
            \mathcal{H} = \Conv \partial\mathcal{C},
        \end{align}
    cf. Eq.~(\ref{Eq:Hull}).
    Since the curvature of $\partial\mathcal{C}$, which shares the sign of expression (\ref{Eq:Curv2}), is non-negative for $d\leq 1/\sqrt{2}$, the boundary of $\mathcal{H}$ is defined in this case by the curve $\partial\mathcal{C}$.

    Summarizing the above considerations, we get that the tight inequalities correspond to the vectors $\boldsymbol{\lambda}$ normal to the boundary curve $\partial\mathcal{C}$.
    Indeed, as it follows from Eq.~(\ref{Eq:BoundSup}), the values of $\alpha$ related to the interior of $\mathcal{C}$ are not assigned to any supporting hyperplane of $\mathcal{H}$.
    This yields $L(\alpha)=\varnothing$ if $\boldsymbol{\Pi}(\alpha)\notin\partial\mathcal{C}$ in Eq.~(\ref{Eq:LambdaSetFormula0}).
    Further careful application of Statement~\ref{St:TightCondition0}, see Appendix~\ref{App:TightUHD}, yields
            \begin{align}\label{Eq:lambdaUHD}
                \boldsymbol{\lambda}(t)=\begin{pmatrix}\dot{\Pi}(0|t;\gamma_2) & -\dot{\Pi}(0|t;\gamma_1)\end{pmatrix}.
            \end{align}
    This implies that the tight inequalities are given by
    		\begin{align}
    			 -(t-d) e^{-(t-d)^2} \mathcal{P}(0|\gamma_1) + (t+d)& e^{-(t+d)^2} \mathcal{P}(0|\gamma_2) \nonumber \\
    			& \leq 2de^{-2( t^2 - d^2)}.
    		\end{align}
	These linear inequalities can be combined into three nonlinear ones as
		\begin{align}
			& \sqrt{-\ln \mathcal{P}(0|\gamma_1)} + \sqrt{-\ln\mathcal{P}(0|\gamma_2)} \geq 2d,\label{Eq:TriangleInequalities1}\\
			& \sqrt{-\ln \mathcal{P}(0|\gamma_1)} + 2d \geq \sqrt{-\ln\mathcal{P}(0|\gamma_2)}, \label{Eq:TriangleInequalities2}\\
			& \sqrt{-\ln\mathcal{P}(0|\gamma_2)} + 2d \geq \sqrt{-\ln \mathcal{P}(0|\gamma_1)},\label{Eq:TriangleInequalities3}
		\end{align}
    cf.~Appendix~\ref{App:Boundary}.
	Violation of at least one of them is a necessary and sufficient condition for nonclassicality of measurement statistics and a sufficient condition for nonclassicality of quantum states.

	We demonstrate the method with squeezed vacuum states $\ket{r}=\hat{S}(r)\ket{0}$.
    This state is supposed to suffer from losses (including detection losses) described by the efficiency $\eta$.
	We consider a challenging scenario by choosing the coherent displacements $\gamma_1$ and $\gamma_2$ to be along the antisqueezed quadrature.
	The points in Fig.~\ref{Fig:UHD} outside the classical region $\mathcal{H}$ correspond to nonclassical probability distributions.
	At the same time, the points inside the classical region $\mathcal{H}$ can be replicated with a coherent state.
	Hence, nonclassicality in such a scheme is in principle testable for high efficiencies.

    Similar to nonclassicality of photocounting statistics for phase-squeezed coherent states, this result seems counterintuitive.
    Indeed, the signal field is displaced by the local oscillator along the antisqueezed quadrature.
    In fact, we observe photocounting statistics of two different phase-squeezed coherent states, whose coherent amplitudes depend on the local-oscillator amplitude.
    As mentioned in Sec.~\ref{Sec:Odd}, nonclassicality of the photocounting statistics for such a state is not an intuitive feature and, moreover, cannot be determined with a single on-off detector.
    Here we show that these states can exhibit nonclassical properties with such a detector.
    However, this interpretation is only possible if we consider this experiment in the context of unbalanced homodyne detection with two device settings.

\section{Summary and conclusions}
\label{Sec:Concl}
	We have considered nonclassicality of measurement statistics in quantum-optical experiments.
    Nonclassicality, in this context, indicates that these statistics cannot be replicated using the same measurement device employed in the original experiment but with classical radiation fields at the source.
    It is worth noting that this property, which is important and interesting itself, is a sufficient condition for a broader concept---nonclassicality of quantum states, i.e., the impossibility of representing them as a statistical mixture of coherent states.

	We have utilized the fact that the classical probability distributions of observables conditioned by device settings belong to the convex hull of such probability distributions for the coherent states.
	Consequently, all nonclassical probability distributions lie outside it.
	Applying the hyperplane separation theorem from convex geometry, one can obtain a set of inequalities whose violation is a necessary and sufficient condition for nonclassicality of measurement statistics.
	In fact, the same mathematical concept is used to test other quantum phenomena such as Bell nonlocality and entanglement.

	The problem is that the set of inequalities, testing nonclassicality of measurement statistics, is not optimal and their application requires the use of involved optimization techniques.
	In this paper, we have shown that this task can be significantly simplified by selecting the optimal set of these inequalities.
	We refer to them as tight inequalities, because they tightly bound the convex set of classical probability distributions.

    First, we applied our method to photocounting statistics.
    In this context, we have considered two types of detectors: ideal PNR and realistic click detectors, i.e., arrays of on-off detectors.
    The simplest nontrivial scenario corresponds to two detectors in the array.
    In this case, we can combine all linear tight inequalities into a single nonlinear one, which gives us a simple necessary and sufficient condition for nonclassicality of the photocounting statistics obtained with such a detection system.

    An important scenario corresponds to the case with three and more detectors in the array.
    In particular, we have derived a subset of tight inequalities for the case of an odd number of such detectors.
    The result is also applicable to a broad class of other detection techniques with an odd maximum number of clicks.
    In particular, we have considered the case of PNR detectors restricted by a given cut-off number of photons.

    We have shown that our inequalities can demonstrate nonclassiacality of photocounting statistics for a counterintuitive example---phase-squeezed coherent states.
    It is well known that such states are characterized by positive excess noise for photocounting statistics and amplitude quadrature.
    Photocounting measurements are typically considered inapplicable to test their nonclassicality.
    We have shown that this common intuition is generally incorrect.
    In fact, photocounting statistics of phase-squeezed coherent states are nonclassical.
    Moreover, our method enables it to be tested with minimal experimental resources.

    Next, we have considered a typical measurement, involving device settings---unbalanced homodyne detection.
    Restricting by use of an on-off detector and two settings of the local oscillators, we have derived a full set of tight inequalities, which test nonclassicality of measurement statistics for this scenario.
    Moreover, we have shown that these linear inequalities are combined in three nonlinear ones.

    The applicability of the method is shown for squeezed vacuum states.
    Similar to the case of photocounting measurements, we have considered a counterintuitive example---the local oscillator displaces this state along the antisqueezed quadrature.
    Intuitively, the corresponding probability distributions should not have nonclassical properties, especially if analyzed with an on-off detector.
    However, we have shown that the probability distributions conditioned by the local-oscillator amplitude for such states are nonclassical.

    Therefore, we have proposed an efficient method for testing nonclassicality of measurement statistics in quantum optics.
    We hope that this result will finds its application in optical experiments testing quantum theory.
    We also believe that it could be useful for a deeper understanding of nonclassical resources in modern quantum technologies.

\begin{acknowledgments}
	The authors acknowledge the National Research Foundation of Ukraine for supporting this work via the Project Nr. 2020.02/0111, Nonclassical and hybrid correlations of quantum systems under realistic conditions.
    A.A.S. thanks Boris Hage for enlightening discussions.
\end{acknowledgments}

\paragraph*{Note added.} Recently, Ref.~\cite{Fiurasek2024} has been published, in which a similar technique has been applied to the unbalanced Hanbury~Brown--Twiss experiment scheme with on-off detectors.

\appendix

\section{Derivation of inequalities for nonclassicality of measurement statistics}
\label{App:DDBell}

    In this section, we remind the reader with the derivation of inequalities~(\ref{Eq:BellLikeInequalities}), introduced in Refs.~{\cite{Semenov2021,Kovtoniuk2022}}.
    Applying Born's rule $\mathcal{P}(A|a)=\Tr[\hat{\Pi}(A|a)\hat{\rho}]$ in Eq.~(\ref{Eq:Pfunction}) we get
		\begin{align}\label{Eq:BornPS}
			\mathcal{P}(A|a)=\int_\mathbb{C}\D^2\alpha\, \Pi\left(A|a;\alpha\right)P(\alpha).
		\end{align}
	Multiplying the above equation by an arbitrary $\lambda(A,a)$ and performing summation over $A$ and $a$, we get
		\begin{align}
			\sum\limits_{a,A}\lambda&(A,a)\mathcal{P}(A|a)\nonumber\\
            &=
			\int_\mathbb{C}\D^2\alpha\,\left[\sum\limits_{a,A}\lambda(a,A) \Pi\left(A|a;\alpha\right)\right]P(\alpha).\label{Eq:DDBell_Der}
		\end{align}
	If $P(\alpha)\geq0$ and $\int_\mathbb{C}\D^2\alpha P(\alpha)=1$, then $\forall f(\alpha)$
		\begin{align}
			\int_\mathbb{C}\D^2\alpha\,f(\alpha)P(\alpha;s)\leq\sup_{\alpha\in\mathbb{C}} f(\alpha)
		\end{align}
    according to the first mean-value theorem.
	Choosing $\textstyle f(\alpha)=\sum\limits_{a,A}\lambda(A,a) \Pi\left(A|a;\alpha\right)$ and using Eq.~(\ref{Eq:DDBell_Der}), we obtain inequalities~(\ref{Eq:BellLikeInequalities}).
    Furthermore, employing the Farkas lemma \cite{Farkas1902}, one can prove that if these inequalities are satisfied, then there exist such $P(\alpha)\geq0$ that Eq.~(\ref{Eq:BornPS}) is satisfied.
    Importantly, this function may not be the same as the $P$ function.
    Consequently, nonclassicality of measurement statistics always implies nonclassicality of quantum states, but the reverse is generally not true.

\section{General statements on tight inequalities}
\label{App:GenSt}

In this section, we provide the general statements used for deriving the tight inequalities and give the detailed proof of Statement~\ref{St:TightCondition0} from Sec.~\ref{Sec:Tight}.

\subsection{Supporting hyperplane for the convex hull of a manifold}
\label{App:GenStHP}

In this section we prove that a supporting hyperplane of $\mathcal{H}$ contains at least one point of $\mathcal{C}$, cf. Eq.~(\ref{Eq:Hull}).
Let $\boldsymbol{\lambda}$ be the normal vector of a supporting hyperplane.
The hyperplane equation corresponds to the equality in inequality (\ref{Eq:VectorBLI}),
    \begin{align}
        \label{Eq:SupportFunction}
        \boldsymbol{\mathcal{P}} \cdot \boldsymbol{\lambda} = \sup_{\alpha \in \mathbb{C}} \boldsymbol{\Pi}(\alpha) \cdot \boldsymbol{\lambda}.
    \end{align}
This implies that the point of the manifold, parametrized by the value of the complex amplitude $\argmax_{\alpha \in \mathbb{C}} \boldsymbol{\Pi}(\alpha) {\cdot} \boldsymbol{\lambda}$, lies within the supporting hyperplane.

\subsection{Closed convex cones}
\label{App:GenStCC}

Let us give a proof of the fact that $L(\alpha)$ defined by Eq.~(\ref{Eq:LambdaSetFormula0}) forms a closed convex cone.
If $\boldsymbol{\lambda}_1$ and $\boldsymbol{\lambda}_2 \in L(\alpha)$, then their conic combination $C_1 \boldsymbol{\lambda}_1 + C_2 \boldsymbol{\lambda}_2$, where $C_1$, $C_2 \geq 0$, also belongs to $L(\alpha)$, since the sum of two functions that share a global maximum point reaches its global maximum at that point.
This proves that $L(\alpha)$ is a convex cone.
In order to prove that it is closed, consider an arbitrary convergent sequence $\boldsymbol{\lambda}_n \in L(\alpha)$, $n \in \mathbb{N}$.
Since for any $n$ and $\alpha^\prime \in \mathbb{C}$
\begin{align}
	\boldsymbol{\Pi}(\alpha) \cdot \boldsymbol{\lambda}_n \geq \boldsymbol{\Pi}(\alpha^\prime) \cdot \boldsymbol{\lambda}_n,
\end{align}
the same property holds for the limit of the sequence
\begin{align}
	\boldsymbol{\Pi}(\alpha) \cdot \boldsymbol{\tilde{\lambda}} \geq \boldsymbol{\Pi}(\alpha^\prime) \cdot \boldsymbol{\tilde{\lambda}},\quad \boldsymbol{\tilde{\lambda}} = \lim_{n \to +\infty} \boldsymbol{\lambda}_n.
\end{align}
Thus, any convergent sequence in $L(\alpha)$ has its limit also in $L(\alpha)$, which proves that $L(\alpha)$ is closed.

\subsection{Proof of Statement~\ref{St:TightCondition0}}
\label{App:GenStSt1}

	Next, we prove Statement~\ref{St:TightCondition0}.
	For this purpose, we first suppose that the set $\Lambda$ is defined by Eq.~(\ref{Eq:LambdaSetFormula}).
	We have to show that the two conditions from Definition~\ref{DefTight} are satisfied for this set.

	Let us start with Condition \ref{Def1-1}.
	Suppose that $\boldsymbol{\lambda^\prime} \not \in \Lambda$ and $\argmax_{\alpha \in \mathbb{C}_{\infty}}  \boldsymbol{\Pi}(\alpha) \cdot\boldsymbol{\lambda^\prime} = \alpha^\prime$.
	We have to prove that the corresponding inequality follows from the inequalities given by the test functions from $\Lambda$.
	The definition of extreme rays yields
	\begin{align}
		\boldsymbol{\lambda^\prime} = \sum\limits_{\boldsymbol{\lambda} \in L(\alpha')} C(\boldsymbol{\lambda}) \boldsymbol{\lambda},
	\end{align}
	where $C(\boldsymbol{\lambda}) \geq 0$.
	This implies that
	\begin{align}
		\boldsymbol{\mathcal{P}} \cdot \boldsymbol{\lambda^\prime}
		&= \sum_{\boldsymbol{\lambda} \in L(\alpha^\prime)} C(\boldsymbol{\lambda}) \boldsymbol{\mathcal{P}} \cdot \boldsymbol{\lambda} \nonumber\\
		&\leq \sum_{\boldsymbol{\lambda} \in L(\alpha^\prime)} C(\boldsymbol{\lambda}) \sup_{\alpha \in \mathbb{C}} \boldsymbol{\Pi}(\alpha) \cdot \boldsymbol{\lambda} \nonumber\\
		&= \sum_{\boldsymbol{\lambda} \in L(\alpha^\prime)} C(\boldsymbol{\lambda}) \boldsymbol{\Pi}(\alpha^\prime) \cdot \boldsymbol{\lambda}
		= \boldsymbol{\Pi}(\alpha^\prime) \cdot \boldsymbol{\lambda^\prime}\nonumber \\
		&= \sup_{\alpha \in \mathbb{C}} \boldsymbol{\Pi}(\alpha) \cdot \boldsymbol{\lambda^\prime}.
	\end{align}
	Thus, the inequality associated with $\boldsymbol{\lambda^\prime}$ follows from those corresponding to test functions from $L(\alpha')$.

	Next, we prove Condition~\ref{Def1-2}.
	Suppose that there is $\boldsymbol{\lambda^\prime} \in \Ext \left[L(\alpha^\prime)\right] \subset \Lambda$ such that the corresponding inequality follows from inequalities associated with some test functions from $\Lambda$.
	Then $\boldsymbol{\lambda^\prime}$ can be represented as a conic combination of test functions from $\Lambda$.
        Let us label these functions as $\boldsymbol{\lambda}_1, \ldots, \boldsymbol{\lambda}_n$,
	\begin{align}
		\boldsymbol{\lambda^\prime} = \sum\limits_{i = 1}^n C_i \boldsymbol{\lambda}_i,\quad C_i > 0,
		\quad \boldsymbol{\lambda}_i \in \Lambda,
	\end{align}
	and
	\begin{align}
		\boldsymbol{\mathcal{P}} \cdot \boldsymbol{\lambda^\prime} = \sum\limits_{i = 1}^n C_i \boldsymbol{\mathcal{P}} \cdot \boldsymbol{\lambda}_i \leq \sum\limits_{i = 1}^n C_i \sup_{\alpha \in \mathbb{C}} \boldsymbol{\Pi}(\alpha) \cdot \boldsymbol{\lambda}_i.
	\end{align}
	At the same time, the right-hand-side of this inequality fulfills the condition
	\begin{align}
		\sum\limits_{i = 1}^n C_i \sup_{\alpha \in \mathbb{C}} \boldsymbol{\Pi}(\alpha) \cdot \boldsymbol{\lambda}_i
		\geq \sup_{\alpha \in \mathbb{C}} \sum\limits_i C_i \boldsymbol{\Pi}(\alpha) \cdot \boldsymbol{\lambda}_i\nonumber\\
		= \sup_{\alpha \in \mathbb{C}} \boldsymbol{\Pi}(\alpha) \cdot \boldsymbol{\lambda^\prime}.
	\end{align}
	Therefore, the inequality for $\boldsymbol{\lambda^\prime}$ can be inferred from the inequalities for $\boldsymbol{\lambda}_i$ only if the equality
	\begin{align}\label{Eq:Sup}
		\sum\limits_{i = 1}^n C_i \sup_{\alpha \in \mathbb{C}} \boldsymbol{\Pi}(\alpha) \cdot \boldsymbol{\lambda}_i = \sup_{\alpha \in \mathbb{C}} \sum\limits_{i = 1}^n C_i \boldsymbol{\Pi}(\alpha) \cdot \boldsymbol{\lambda}_i
	\end{align}
	is satisfied.
	This implies
	\begin{align}\label{Eq:SupAlphaPr}
		\sup_{\alpha \in \mathbb{C}} \boldsymbol{\Pi}(\alpha) \cdot \boldsymbol{\lambda}_i = \boldsymbol{\Pi}(\alpha^\prime) \cdot \boldsymbol{\lambda}_i
	\end{align}
	for all $i =1\ldots n$.

	Indeed, let us assume that Eq.~(\ref{Eq:Sup}) is satisfied and, without loss of generality, Eq.~(\ref{Eq:SupAlphaPr}) is not satisfied for $\boldsymbol{\lambda}_j$, $j = 1 \ldots k \leq n$.
	Hence, these $\boldsymbol{\lambda}_j$ fulfill the condition
	\begin{align}
		\sup_{\alpha \in \mathbb{C}} \boldsymbol{\Pi}(\alpha) \cdot \boldsymbol{\lambda}_j = \boldsymbol{\Pi}(\alpha_j) \cdot \boldsymbol{\lambda}_j > \boldsymbol{\Pi}(\alpha^\prime) \cdot \boldsymbol{\lambda}_j.
	\end{align}
	This yields
	\begin{align}
		&\sum\limits_{i = 1}^n \sup_{\alpha \in \mathbb{C}} C_i \boldsymbol{\Pi}(\alpha) \cdot \boldsymbol{\lambda}_i \nonumber\\
		&= \sum\limits_{j=1}^k C_j \boldsymbol{\Pi}(\alpha_j) \cdot \boldsymbol{\lambda}_j + \sum_{i = k + 1}^n C_i \boldsymbol{\Pi}(\alpha^\prime) \cdot \boldsymbol{\lambda}_i.
	\end{align}
	On the other hand, Eq.~(\ref{Eq:Sup}) implies
	\begin{align}
		&\sum\limits_{i = 1}^n \sup_{\alpha \in \mathbb{C}} C_i \boldsymbol{\Pi}(\alpha) \cdot \boldsymbol{\lambda}_i 
		= \sum\limits_{i=1}^n C_i \boldsymbol{\Pi}(\alpha^\prime) \cdot \boldsymbol{\lambda}_i.
	\end{align}
	Equating these expressions we obtain the equality
	\begin{align}
		\sum\limits_{j = 1}^k C_j \boldsymbol{\Pi}(\alpha_j) \cdot \boldsymbol{\lambda}_j = \sum_{j = 1}^k C_j \boldsymbol{\Pi}(\alpha') \cdot \boldsymbol{\lambda}_j.
	\end{align}
	However, it contradicts the assumption $\boldsymbol{\Pi}(\alpha_j) \cdot \boldsymbol{\lambda}_j > \boldsymbol{\Pi}(\alpha') \cdot \boldsymbol{\lambda}_j$.
	This implies
	\begin{align}
		\sup_{\alpha \in \mathbb{C}} \boldsymbol{\Pi}(\alpha) \cdot \boldsymbol{\lambda}_i = \boldsymbol{\Pi}(\alpha^\prime) \cdot \boldsymbol{\lambda}_i,
	\end{align}
	where $i = 1 \ldots n$, that is $\boldsymbol{\lambda}_i \in L(\alpha^\prime)$.

	This means that $\boldsymbol{\lambda}^\prime$ is a conic combination of other test functions from $\Ext \left[ L(\alpha^\prime) \right]$.
	This conclusion violates the assumption that $\boldsymbol{\lambda}^\prime$ belongs to $\Ext \left[ L(\alpha^\prime) \right]$.
	Thus, the inequality for $\boldsymbol{\lambda}^\prime$ cannot be inferred from other tight inequalities.

\section{Nonlinear inequality for photocounting statistics with click detectors in the case of $N=2$}
\label{App:N2}
	In this section we sketch the derivation of inequality (\ref{Eq:NonlinearPD}) from Sec.~\ref{Sec:PC2}.
	Applying $N=2$ and $t=e^{-|\alpha|^2/2}$ to Eq.~(\ref{Eq:ArrayDetectorsPOVM}), we get
	\begin{align}
		&\Pi(0|t) =  t^2,\label{Eq:ParN2-0}\\
		&\Pi(1|t) = 2 t \left( 1 - t \right).\label{Eq:ParN2-1}
	\end{align}
	From Eq.~(\ref{Eq:ParN2-0}) we get $t=\sqrt{\Pi(0|t)}$.
	Substituting it into Eq.~(\ref{Eq:ParN2-1}), we arrive at
	\begin{align}
		\Pi(1|t) =2 \left[\sqrt{\Pi(0|t)}-\Pi(0|t)\right].
	\end{align}
	This implies that the function $\Pi_1(\mathcal{P}(0))$, cf. Eq.~(\ref{Eq:Curve}), is given by
	\begin{align}
		\Pi_1(\mathcal{P}(0))=2 \left[\sqrt{\mathcal{P}(0)}-\mathcal{P}(0)\right].
	\end{align}
	Applying it in inequality (\ref{Eq:IneqDetectors}) leads to
	\begin{align}
		\mathcal{P}(1)+2\mathcal{P}(0)\leq2\sqrt{\mathcal{P}(0)},
	\end{align}
	which is equal to inequality (\ref{Eq:NonlinearPD}).

\section{Nonlinear inequalities for photocounting statistics in the case of $N=3$}
\label{App:N3}

    In this section we present the derivation of nonlinear inequalities for photocounting statistics in the case of $N=3$ from Sec.~\ref{Sec:PC3} given by Eqs.~(\ref{Eq:PNR3_1}) and (\ref{Eq:PNR3_2}) for the PNR detectors and by Eqs.~(\ref{Eq:Array3_1}) and (\ref{Eq:Array3_2}) for the click detectors.
    Let us start with the PNR detectors.
    The tight inequalities with the test function $\boldsymbol{\lambda}(t_1;\tau)$, cf. Eq.~(\ref{Eq:TightLambda3D}), are given by
        \begin{align}\label{Eq:TightIneqPNR3_1}
            -\mathcal{P}(0)\frac{u^2}{2}+\mathcal{P}(1)u-\mathcal{P}(2)\leq0
        \end{align}
    for $\tau=0$ and
        \begin{align}
           \mathcal{P}(0)\frac{u^2}{2}e^{-u} &-\mathcal{P}(1)\frac{u}{2}\left(-2+2e^{-u}+u\right)\nonumber\\
           &+\mathcal{P}(2)\left(-1+e^{-u}+u\right)\leq \frac{u^2}{2}e^{-u}\label{Eq:TightIneqPNR3_2}
        \end{align}
    for $\tau=1$.
    Here $u=|\alpha|^2=-3\ln t_1\in[0,+\infty)$.

    Inequality (\ref{Eq:TightIneqPNR3_1}) holds if the square trinomial on its left-hand side has either one root or no roots.
    This is the case if its discriminant is nonpositive.
    This condition leads to inequality (\ref{Eq:PNR3_1}), which is a Klyshko inequality.

    Let us now multiply inequality (\ref{Eq:TightIneqPNR3_2}) by the non-negative function $e^{u}/u^2$ and write it as
               \begin{align}
           \frac{\mathcal{P}(0)-1}{2} -\frac{\mathcal{P}(1)}{2u}&\left(-2e^{u}+2+ue^{u}\right)\nonumber\\
           +\mathcal{P}(2)&\left(-e^{u}+1+ue^{u}\right)\leq 0.\label{Eq:TightIneqPNR3_2_M}
        \end{align}
    The left-hand side of this inequality has the maximum at $u=2\mathcal{P}(2)/\mathcal{P}(1)$.
    Substituting this value into inequality (\ref{Eq:TightIneqPNR3_2}) we get inequality (\ref{Eq:PNR3_2}).

    Let us now consider the click detectors.
    In this case, the tight inequalities with the test function $\boldsymbol{\lambda}(t_1;\tau)$, cf. Eq.~(\ref{Eq:TightLambda3D}), read
        \begin{align}\label{Eq:TightIneqArray3_1}
            -3(1 - t_1)^2 \mathcal{P}(0) + 2t_1(1 - t_1) \mathcal{P}(1) - t_1^2 \mathcal{P}(2) \leq 0
        \end{align}
    for $\tau=0$ and
        \begin{align}\label{Eq:TightIneqArray3_2}
            3t_1^2 \mathcal{P}(0) + (2t_1^2 + 2t_1 - 1) &\mathcal{P}(1)\nonumber\\
            & + t_1(t_1 + 2) \mathcal{P}(2) - 3t_1^2 \leq 0
        \end{align}
    for $\tau=1$.
    Similarly to the case of PNR detectors, these are square trinomials and the inequalities hold if they have either one root or no roots.
    By enforcing these conditions we arrive at inequalities (\ref{Eq:Array3_1}) and (\ref{Eq:Array3_2}).

\section{Tight inequalities for photocounting with odd $N$}
\label{App:Odd}

In this section, we give details related to Sec.~\ref{Sec:Odd}.
In particular, we recall the definition of the Hodge star, provide the proof of Statement~\ref{St:PhotoCount1}, and demonstrate that the $Q$ symbols of the POVM for the PNR and the click detectors satisfy the conditions of Statement~\ref{St:PhotoCount1}.

\subsection{Hodge star}
\label{App:Hodge}
Let us first elaborate on the usage of the Hodge star $\star$ and the wedge symbol $\wedge$.
For our purposes, we only need to evaluate the expression $\boldsymbol{u} = \star \bigwedge_{i = 1}^{n - 1} \boldsymbol{v}^i$, where $\boldsymbol{v}^i \in \mathbb{R}^n$.
In such a case $\boldsymbol{u}$ is also a vector from $\mathbb{R}^n$ given by
\begin{align}\label{Eq:Hodge}
	\star \bigwedge_{i = 1}^{n-1} \boldsymbol{v}^i =
	\begin{vmatrix}
		\boldsymbol{e}^1 & \boldsymbol{e}^2 & \dots & \boldsymbol{e}^n \\
		v^1_1 & v^1_2 & \dots & v^1_n \\
		\vdots & \vdots & \ddots & \vdots \\
		v^{n-1}_1 & v^{n-1}_2 & \dots & v^{n-1}_n \\
	\end{vmatrix},
\end{align}
where $\left\{\boldsymbol{e}^1, \boldsymbol{e}^2, \ldots, \boldsymbol{e}^n \right\}$ is the standard basis in $\mathbb{R}^n$, i.e., $e^i_j = \delta_{ij}$.
In other words, $\boldsymbol{u}$ is a generalized cross product of $n - 1$ vectors in an $n$-dimensional space.

\subsection{Proof of Statement~\ref{St:PhotoCount1}}
\label{App:Prof-St2}
	Without loss of generality, let us assume that $\tau = 0$.
	Equation~(\ref{Eq:Cond1}) together with Rolle's theorem implies that $\dot{\boldsymbol{\Pi}}(t) \cdot \boldsymbol{\lambda}(\boldsymbol{t};\tau)$ equals zero at the points $c_1 \in (0, t_1)$, $c_2 \in (t_1, t_2)$, $\ldots$, $c_m \in (t_{m - 1}, t_m)$; cf.~Fig.~\ref{Fig:CrPoints}.
	Together with $t_1, \ldots, t_m$, these points exhaust the $2m$ zeros of $\dot{\boldsymbol{\Pi}}(t) \cdot \boldsymbol{\lambda}(\boldsymbol{t};\tau)$ as a function of $t$.

	Since the derivatives $\dot{\boldsymbol{\Pi}}(t) \cdot \boldsymbol{\lambda}(\boldsymbol{t};\tau)$ are equal for all $t=t_i$ and $t=c_i$, the $2m - 1$ zeros of $\ddot{\boldsymbol{\Pi}}(t) \cdot \boldsymbol{\lambda}(\boldsymbol{t};\tau)$ are located at $\left(c_1, t_1\right)$, $\ldots$, $\left(c_m, t_m \right)$ according to Rolle's theorem.
	Therefore, we get $\ddot{\boldsymbol{\Pi}}(t_i) \cdot \boldsymbol{\lambda}(\boldsymbol{t};\tau)\neq 0$ for $i=1\ldots m$.
	This implies that the points $t_1, \ldots, t_m$, $c_1, \ldots, c_m$ are extrema of $\boldsymbol{\Pi}(t) \cdot \boldsymbol{\lambda}(\boldsymbol{t};\tau)$.

	Let us set such a sign of $\boldsymbol{\lambda}(\boldsymbol{t};\tau)$ that $t=\tau=0$ is a local maximum of $\boldsymbol{\Pi}(t) \cdot \boldsymbol{\lambda}(\boldsymbol{t};\tau)$ at $[0, 1]$.
	The nearest extremum is $c_1$, which consequently must be a local minimum.
	It yields that $t_1$ is a local maximum.
	This pattern is repeated up until $t_m$, which proves that $\tau, t_1, \ldots, t_m$ are maxima points.
	By conditions of this Statement there are no other critical points at $(0, 1)$.
	Consequently, there are no other points, which could be the global maximum.
	Finally, one has to consider the other endpoint $t = 1 - \tau = 1$.
	Since $t_m$ is a local maximum and there are no extrema in $(t_m,1{-}\tau)$, the point $t=1 - \tau$ is a local minimum.

\subsection{Conditions of Statement~\ref{St:PhotoCount1} for PNR and click detectors}
\label{App:CondDet}

Let us prove that this condition is satisfied by the photon-number resolving (PNR) and click detectors.
We start with the PNR detectors.
For convenience, we use here another parametrization, $s=e^{-|\alpha|^2}$, related to our standard parametrization, $t=e^{-|\alpha|^2/N}$, via the monotonic function $s=t^N$.
In this case, the $Q$ symbols of the POVM are given by
\begin{align}
	\Pi(n|s) = s \frac{(-\ln s)^n}{n!}.
\end{align}
This yields,
\begin{align}
	\boldsymbol{\Pi}(s) \cdot \boldsymbol{\lambda}(\boldsymbol{s};\tau)=s P(-\ln s),
\end{align}
where $P(-\ln s)$ is a $2m$-th degree polynomial in terms of $-\ln s$.
The first derivative,
\begin{align}
	\dot{\boldsymbol{\Pi}}(s) \cdot \boldsymbol{\lambda}(\boldsymbol{s};\tau) = P(-\ln s) - P^\prime(-\ln s),
\end{align}
has $2m$ zeros.
Therefore, apart from $s_1,\ldots s_m$, the function $\boldsymbol{\Pi}(s) \cdot \boldsymbol{\lambda}(\boldsymbol{s}, \tau)$ has $m$ other critical points at $s \in (0, 1)$.
The second derivative reads
\begin{align}
	\ddot{\boldsymbol{\Pi}}(s) \cdot \boldsymbol{\lambda}(\boldsymbol{s};\tau)
	= \frac{P^{\prime\prime}(-\ln s) - P^{\prime}(-\ln s)}{s}.
\end{align}
It has no more than $2m - 1$ zeros, since the nominator in this equation is a $(2m - 1)$-th degree polynomial.
Since the parameters $s$ and $t$ are related through a monotonic function, these conclusions also holds for the latter.

Next, we consider the click detectors with the $Q$ symbols of the POVM given by
\begin{align}
	\Pi(n|t) = \binom{N}{n} t^{N - n} \left( 1 - t \right)^n.
\end{align}
Then $\boldsymbol{\Pi}(t) \cdot \boldsymbol{\lambda}(\boldsymbol{t};\tau)$ is a $(2m + 1)$-th degree polynomial, which has no more than $2m$ critical points and whose second derivative has no more than $2m - 1$ zeros.

\subsection{Extreme-ray condition}
\label{App:ExtrRays}

In Sec.~\ref{Sec:Odd} we have applied a statement from the convex analysis, which can be formulated based on results presented in Ref.~\cite{Rockafellar_book}.
For convenience, we also present it here in notations of this paper.

\begin{statement}
	\label{St:TightCondition}
	Suppose a vector $\boldsymbol{\lambda} \in L$.
	Then it belongs to $\Ext \left[L\right]$ if and only if it is not a conic combination of any other two vectors $\boldsymbol{\lambda}_1$, $\boldsymbol{\lambda}_2 \in L$, i.e., $\boldsymbol{\lambda} \neq C_1 \boldsymbol{\lambda}_1 + C_2 \boldsymbol{\lambda}_2$, $C_1, C_2 > 0$, where $\boldsymbol{\lambda}_i \neq b_i \boldsymbol{\lambda}$, $b_i > 0$, $i = 1, 2$.
\end{statement}
\begin{proof}
	Suppose that there exists $\boldsymbol{\lambda} \in \Ext L$ such that
	\begin{align}
		\boldsymbol{\lambda} = C_1 \boldsymbol{\lambda}_1 + C_2 \boldsymbol{\lambda}_2, \quad c_1, c_2 > 0,
	\end{align}
	for some $\boldsymbol{\lambda}_1$, $\boldsymbol{\lambda}_2$, where $\boldsymbol{\lambda}_i \neq b_i \boldsymbol{\lambda}$, $b_i > 0$, $i = 1, 2$.
	Utilizing
	\begin{align}
		\boldsymbol{\lambda}_i = \sum\limits_{\boldsymbol{\lambda^\prime} \in \Ext [L]} a_i(\boldsymbol{\lambda^\prime}) \boldsymbol{\lambda^\prime},
	\end{align}
	where $a_i(\boldsymbol{\lambda'}) \geq 0$, $i = 1, 2$, one can obtain
	\begin{align}
		\boldsymbol{\lambda} = \sum\limits_{\boldsymbol{\lambda}^\prime \in \Ext [L]} \left(C_1 a_1(\boldsymbol{\lambda^\prime}) + C_2 a_2(\boldsymbol{\lambda}^\prime)\right) \boldsymbol{\lambda}^\prime.
	\end{align}
	Since $\boldsymbol{\lambda} \in  \Ext [L]$, $C_1 a_1(\boldsymbol{\lambda}^\prime) + C_2 a_2(\boldsymbol{\lambda}^\prime) = 0$ for $\boldsymbol{\lambda'} \neq \boldsymbol{\lambda}$.
	Then $a_1(\boldsymbol{\lambda}') = a_2(\boldsymbol{\lambda}') = 0$ if $\boldsymbol{\lambda'} \neq \boldsymbol{\lambda}$.
	But this yields $\boldsymbol{\lambda}_i = a_i(\boldsymbol{\lambda}) \boldsymbol{\lambda}$, which contradicts the initial assumption.

	To prove the converse statement, recall that any test function from $L$ can be expressed as a conic combination of the test functions from $\Ext [L]$.
	But the only possible conic combination representation of $\boldsymbol{\lambda}$ is $\boldsymbol{\lambda}$ itself.
	Therefore, $\boldsymbol{\lambda} \in \Ext [L]$.
\end{proof}

\section{Tight inequalities for unbalanced homodyne detection }
\label{App:UHD}

In this section, we provide details of the derivation of the tight inequalities for the case of unbalanced homodyne detection considered in Sec.~\ref{Sec:UHD}.

\subsection{Mode mismatch}
\label{App:ModeMismatch}

To consider the effect of mode mismatch, we remind the corresponding $Q$ symbols of the POVM, cf. Ref.~\cite{banaszek2002},
\begin{align}
	\Pi(0|\alpha;\gamma;\eta;\xi) = e^{-\eta|\alpha - \gamma|^2} e^{-|\gamma|^2 \eta \frac{1 - \xi}{\xi}},
\end{align}
where the mode mismatch parameter $\xi \in \left[0, 1\right]$ and the detection efficiency $\eta \in \left[0, 1\right]$.
By defining a new parameter $\beta = \sqrt{\eta} \alpha$ it can be rewritten as
\begin{align}
	\Pi(0|\beta;\gamma;\eta;\xi) = e^{-|\beta - \sqrt{\eta}{\gamma}|^2} e^{-|\sqrt{\eta} \gamma|^2 \frac{1 - \xi}{\xi}}.
\end{align}
This is effectively the $Q$ symbols of the POVM with the coherent displacement $\sqrt{\eta} \gamma$ multiplied by the factor
\begin{align}
	g(\gamma;\eta;\xi) = e^{-|\sqrt{\eta} \gamma|^2 \frac{1 - \xi}{\xi}}
\end{align}
Let us find the tight inequalities for the scenario with coherent displacements $\gamma_1$, $\gamma_2$, the efficiency $\eta$, and the mode mismatch parameter $\xi$.
Since the coherent displacements are effectively $\sqrt{\eta} \gamma_1$ and $\sqrt{\eta} \gamma_2$, we can make use of the tight test functions for the case with these settings.
To account for the factor
\begin{align}
	e^{-|\sqrt{\eta} \gamma|^2 \frac{1 - \xi}{\xi}},
\end{align}
we can multiply the corresponding elements of the test functions by the inverse factors.
More specifically, a test function $\boldsymbol{\lambda} = \left\{ \lambda_1, \lambda_2 \right\}$ becomes
\begin{align}
	\boldsymbol{\lambda^\prime} = \left\{ \frac{1}{g(\gamma_1;\eta;\xi)} \lambda_1, \frac{1}{g(\gamma_2;\eta;\xi)} \lambda_2 \right\}.
\end{align}
Therefore, the right-hand side of such inequalities reads
\begin{align}
	& \Pi(0|\beta;\gamma_1;\eta;\xi) \lambda_1^\prime + \Pi(0|\beta;\gamma_2;\eta;\xi) \lambda_2^\prime = \nonumber \\
	= & \Pi(0|\beta;\sqrt{\eta}\gamma_1;1;1) \lambda_1 + \Pi(0|\beta;\sqrt{\eta}\gamma_2;1;1) \lambda_2,
\end{align}
which have been proved to be tight.

\subsection{Boundary curve of $\mathcal{C}$}
\label{App:Boundary}

\begin{figure}[t!]
	\centering
	\includegraphics{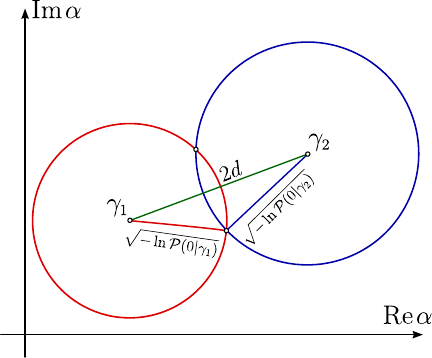}
	\caption{\label{Fig:Circles} The circles, corresponding to Eq.~(\ref{Eq:Circles}), and the triangle, demonstrating inequalities~(\ref{Eq:TriangleInequalities1})--(\ref{Eq:TriangleInequalities3}) are shown.}
\end{figure}

Let us calculate the boundary curve $\partial\mathcal{C}$ of the manifold $\mathcal{C}$.
Suppose that $\boldsymbol{\mathcal{P}} \in \mathcal{C}$, where $\boldsymbol{\mathcal{P}}=\begin{pmatrix} \mathcal{P}(0|\gamma_1) & \mathcal{P}(0|\gamma_2)\end{pmatrix}$. Then
\begin{align}
	\mathcal{P}(0|\gamma_i)= \exp\left(-\left|\alpha - \gamma_i\right|^2\right),
\end{align}
which implies
\begin{align}\label{Eq:Circles}
	\left|\alpha - \gamma_i\right|^2 = -\ln{\mathcal{P}(0|\gamma_i)},\quad i = 1, 2.
\end{align}
These two equations correspond to the circles in the $\alpha$ plane with radii $\sqrt{-\ln{\mathcal{P}(0|\gamma_1)}}$ and $\sqrt{-\ln{\mathcal{P}(0|\gamma_2)}}$, cf.~Fig.~\ref{Fig:Circles}.
If this system of equations has a solution $\alpha$, the two circles have at least one intersection point.
Such an intersection point and the two centers form a triangle with the sides that equal $\sqrt{-\ln{\mathcal{P}(0|\gamma_1)}}$, $\sqrt{-\ln{\mathcal{P}(0|\gamma_2)}}$ and $2d=|\gamma_1{-}\gamma_2|$, which yields the triangle inequalities~(\ref{Eq:TriangleInequalities1})--(\ref{Eq:TriangleInequalities3}).

The inequalities are nonstrict, since if the triangle degenerates into a line, the two circles still have an intersection point.
In this case, the probability values belong to the boundary $\partial\mathcal{C}$.
The latter can be parametrized by Eqs.~(\ref{Eq:UHD-Bound-1}) and (\ref{Eq:UHD-Bound-2}).
Indeed, by substituting this parametrization in the triangle inequalities, we find that inequalities (\ref{Eq:TriangleInequalities2}), (\ref{Eq:TriangleInequalities1}), and (\ref{Eq:TriangleInequalities3}) become equalities for $t\leq-d$, $|t|\leq d$, and $t\geq d$, respectively.
Two other inequalities are trivially satisfied in these cases.
This yields Eq.~(\ref{Eq:BoundSup}) and $L(t) \equiv L(\boldsymbol{\Pi}(t))$.
Here the components of $\boldsymbol{\Pi}(t)$ are given by Eqs.~(\ref{Eq:UHD-Bound-1}) and (\ref{Eq:UHD-Bound-2}).

\subsection{Boundary supremum}
\label{App:BounSup}

Consider the right-hand side of inequality (\ref{Eq:VectorBLI}) in the case of unbalanced homodyne detection with two settings.
To calculate the supremum, let us make the change of variables $\boldsymbol{w} = \boldsymbol{\Pi}(\alpha)$ so that
\begin{align}
	\sup_{\alpha \in \mathbb{C}} \boldsymbol{\Pi}(\alpha) \cdot \boldsymbol{\lambda} = \sup_{\boldsymbol{w} \in \mathcal{C}} \boldsymbol{w} \cdot \boldsymbol{\lambda},
\end{align}
where $\mathcal{C}$ is given by Eq.~(\ref{Eq:Manifold}).
Then $\partial_{w_i} \left(\boldsymbol{w} \cdot \boldsymbol{\lambda} \right) = \lambda_i$ for $i = 1, 2$.
These partial derivatives equal zero simultaneously if and only if $\lambda_i = 0$, $i = 1, 2$, which is a trivial case.
Otherwise,
\begin{align}
	\sup_{\boldsymbol{w} \in \mathcal{C}} \boldsymbol{w} \cdot \boldsymbol{\lambda} = \sup_{\boldsymbol{w} \in \partial\mathcal{C}} \boldsymbol{w} \cdot \boldsymbol{\lambda},
\end{align}
i.e., the maximum value of $\boldsymbol{w} \cdot \boldsymbol{\lambda}$ is reached at the boundary $\partial\mathcal{C}$ of $\mathcal{C}$.

\subsection{Derivation of tight inequalities}
\label{App:TightUHD}

Let us consider properties of the boundary curve $\partial\mathcal{C}$.
Let $\theta(t)$ be the tangential angle of the curve at the point $\boldsymbol{\Pi}(t)$, i.e., the polar angle of
\begin{align}
	\dot{\boldsymbol{\Pi}}(t) = -2\begin{pmatrix}[t + d] \Pi(0|t;\gamma_1)&
		[t - d]\Pi(0|t;\gamma_2)\end{pmatrix}.
\end{align}
First, note that
\begin{align}
	\frac{\dot{\boldsymbol{\Pi}}(-\infty)}{\left| \dot{\boldsymbol{\Pi}}(-\infty) \right|}  =
	\begin{pmatrix}
		1 & 0
	\end{pmatrix}
	\quad\textrm{and}\quad
	\frac{\dot{\boldsymbol{\Pi}}(+\infty)}{\left| \dot{\boldsymbol{\Pi}}(+\infty) \right|}  =
	\begin{pmatrix}
		0 & -1
	\end{pmatrix}.
\end{align}
This implies $\theta(-\infty) = 0$ and $\theta(+\infty) = 3\pi/2$.
Second, since $\Pi(0|t;\gamma_1), \Pi(0|t;\gamma_2) \geq 0$, it is evident that the conditions $\dot{\Pi}(0|t;\gamma_1) > 0$, $\dot{\Pi}(0|t;\gamma_2) < 0$ can never hold simultaneously.
This implies that $0 \leq \theta(t) \leq 3\pi/2$.
Third, the tangential angle can be represented as
\begin{align}
	\label{Eq:TangentialAngle}
	&\theta(t) = \\
	&\int\limits_{-\infty}^{t}\frac{ \dot{\Pi}(0|t^\prime;\gamma_1) \ddot{\Pi}(0|t^\prime;\gamma_2) {-} \dot{\Pi}(0|t^\prime;\gamma_2) \ddot{\Pi}(0|t^\prime;\gamma_1) } {\left| \dot{\boldsymbol{\Pi}}(t') \right|^2} d t^\prime.\nonumber
\end{align}
Consider the quantity
\begin{align}
	\dot{\Pi}&(0|t;\gamma_1) \ddot{\Pi}(0|t;\gamma_2) - \dot{\Pi}(0|t;\gamma_2) \ddot{\Pi}(0|t;\gamma_1) \label{Eq:Curv2}\\
	&= 16 \Pi(0|t;\gamma_1) \Pi(0|t;\gamma_2) d \left( t^2 - d^2 + \frac{1}{2} \right).\nonumber
\end{align}
If $d \leq 1/\sqrt{2}$, this expression is non-negative, and hence $\dot{\theta}(t) \geq 0$.
In other words, $\theta(t)$ is a monotonic function in this case.

Let us calculate $L(t)$ for $t \in \mathbb{R}$.
We focus on the case when $d \leq 1/\sqrt{2}$.
If
\begin{align}
	\argmax\limits_{t' \in \mathbb{R}_{\infty}}\boldsymbol{\Pi}(t') \cdot \boldsymbol{\lambda} = t,
\end{align}
then
\begin{align}
	\frac{\partial}{\partial t'} \left[\boldsymbol{\Pi}(t') \cdot \boldsymbol{\lambda}(t) \right]_{t' = t} = \dot{\boldsymbol{\Pi}}(t) \cdot \boldsymbol{\lambda}(t) = 0.
\end{align}
This yields
\begin{align}
	\boldsymbol{\lambda}(t)=C\begin{pmatrix}\dot{\Pi}(0|t;\gamma_2) & -\dot{\Pi}(0|t;\gamma_1)\end{pmatrix},
\end{align}
$C \in \mathbb{R}$.
Consider the second derivative
\begin{align}
	\label{Eq:UHDSecondDerivative}
	&\frac{\partial^2}{\partial t^{\prime2}}\big[\boldsymbol{\Pi}(t^\prime)\cdot \boldsymbol{\lambda}(t)\big]_{t^\prime=t}=\ddot{\boldsymbol{\Pi}}(t) \cdot \boldsymbol{\lambda}(t)\nonumber\\
	&= C \left[\dot{\Pi}(0|t;\gamma_2) \ddot{\Pi}(0|t;\gamma_1)-\dot{\Pi}(0|t;\gamma_1) \ddot{\Pi}(0|t;\gamma_2) \right]\nonumber\\
	&\propto C \dot{\theta}(t).
\end{align}
The expression inside the brackets is positive for $d < 1/\sqrt{2}$, cf.~Eq.~(\ref{Eq:Curv2}).
Hence, the sign of the second derivative is determined by the sign of $C$.
For $t$ to be a local maximum of $\boldsymbol{\Pi}(t^\prime) \cdot \boldsymbol{\lambda}(t)$ as a function of $t^\prime$, it is sufficient that the second derivative is negative, and hence $C > 0$.

If $d{=}1/\sqrt{2}$, the second derivative is zero, therefore we should consider this case separately.
For $t \neq 0$, Eq.~(\ref{Eq:UHDSecondDerivative}) is still negative for $C > 0$.
If $t = 0$, one directly can check that
\begin{align}
	\frac{\partial^3}{\partial t^{\prime 3}} \left[ \boldsymbol{\Pi}(t^\prime) \cdot \boldsymbol{\lambda}(0) \right]_{t^\prime = 0} = 0
\end{align}
and
\begin{align}
	\frac{\partial^4}{\partial t^{\prime 4}} \left[ \boldsymbol{\Pi}(t^\prime) \cdot \boldsymbol{\lambda}(0) \right]_{t^\prime = 0} = -C \frac{16\sqrt{2}}{e},
\end{align}
which is negative if $C > 0$.
This ensures that $t^\prime = 0$ is a local maximum of $\boldsymbol{\Pi}(t^\prime) \cdot \boldsymbol{\lambda}(0)$.

Suppose that there is another extremum of $\boldsymbol{\Pi}(t^\prime) \cdot \boldsymbol{\lambda}(t)$ as a function of $t^\prime$ at the point $t^\prime=s\neq t$.
The condition
\begin{align}
	\frac{\partial}{\partial t^\prime} \left[\boldsymbol{\Pi}(t^\prime) \cdot \boldsymbol{\lambda}(t) \right]_{t^\prime=s} = \dot{\boldsymbol{\Pi}}(s)\cdot \boldsymbol{\lambda}(t) = 0
\end{align}
requires that $\dot{\boldsymbol{\Pi}}(s) = B \dot{\boldsymbol{\Pi}}(t)$, $B \neq 0$ since both $\dot{\boldsymbol{\Pi}}(s)$ and $\dot{\boldsymbol{\Pi}}(t)$ are orthogonal to $\boldsymbol{\lambda}(t)$.
The condition $s\neq t$ implies that $B < 0$.
Indeed, $B > 0$ yields $\theta(s) = \theta(t)$ since the vectors $\dot{\boldsymbol{\Pi}}(t)$ and $\dot{\boldsymbol{\Pi}}(s)$ are co-directed.
Hence, $s = t$, since $\theta(t^\prime)$ is a monotonic function.
The second derivative at this point is positive, i.e.,
\begin{align}
	\frac{\partial^2}{\partial t'^2} \left[\boldsymbol{\Pi}(t') \cdot \boldsymbol{\lambda}(t) \right]_{t^\prime =s} = \ddot{\boldsymbol{\Pi}}(s) \cdot \boldsymbol{\lambda}(t)\nonumber\\
	= B \ddot{\boldsymbol{\Pi}}(t) \cdot \boldsymbol{\lambda}(t) > 0,
\end{align}
indicating that $t^\prime=s$ is the point of a local minimum.

Finally, in order to prove that $t' = t$ corresponds to the global maximum of $\boldsymbol{\Pi}(t') \cdot \boldsymbol{\lambda}(t)$, we need to consider the infinite point $t = \infty$.
We employ the property
\begin{align}
	\boldsymbol{\Pi}(\infty) \cdot \boldsymbol{\lambda}(t)= 0 < \boldsymbol{\Pi}(t) \cdot \boldsymbol{\lambda}(t) = 4C e^{2t^2 + 2d^2} d.
\end{align}
This implies that
\begin{align}
	L(t) = \left\{ C \begin{pmatrix}
		\dot{\Pi}_2(t) & -\dot{\Pi}_1(t)
	\end{pmatrix} \middle| C > 0 \right\}.
\end{align}
Therefore, the extreme rays of $L(t)$ are given by
\begin{align}
	\Ext L(t) = \left\{ \begin{pmatrix}
		\dot{\Pi}_2(t) & -\dot{\Pi}_1(t)
	\end{pmatrix}  \right\},
\end{align}
which defines tight inequalities.
According to Statement~\ref{St:TightCondition0}, the set of tight inequalities is given by the vectors $\boldsymbol{\lambda}$ defined by Eq.~(\ref{Eq:lambdaUHD}).

\bibliography{biblio}

\begin{thebibliography}{76}%
\makeatletter
\providecommand \@ifxundefined [1]{%
 \@ifx{#1\undefined}
}%
\providecommand \@ifnum [1]{%
 \ifnum #1\expandafter \@firstoftwo
 \else \expandafter \@secondoftwo
 \fi
}%
\providecommand \@ifx [1]{%
 \ifx #1\expandafter \@firstoftwo
 \else \expandafter \@secondoftwo
 \fi
}%
\providecommand \natexlab [1]{#1}%
\providecommand \enquote  [1]{``#1''}%
\providecommand \bibnamefont  [1]{#1}%
\providecommand \bibfnamefont [1]{#1}%
\providecommand \citenamefont [1]{#1}%
\providecommand \href@noop [0]{\@secondoftwo}%
\providecommand \href [0]{\begingroup \@sanitize@url \@href}%
\providecommand \@href[1]{\@@startlink{#1}\@@href}%
\providecommand \@@href[1]{\endgroup#1\@@endlink}%
\providecommand \@sanitize@url [0]{\catcode `\\12\catcode `\$12\catcode
  `\&12\catcode `\#12\catcode `\^12\catcode `\_12\catcode `\%12\relax}%
\providecommand \@@startlink[1]{}%
\providecommand \@@endlink[0]{}%
\providecommand \url  [0]{\begingroup\@sanitize@url \@url }%
\providecommand \@url [1]{\endgroup\@href {#1}{\urlprefix }}%
\providecommand \urlprefix  [0]{URL }%
\providecommand \Eprint [0]{\href }%
\providecommand \doibase [0]{https://doi.org/}%
\providecommand \selectlanguage [0]{\@gobble}%
\providecommand \bibinfo  [0]{\@secondoftwo}%
\providecommand \bibfield  [0]{\@secondoftwo}%
\providecommand \translation [1]{[#1]}%
\providecommand \BibitemOpen [0]{}%
\providecommand \bibitemStop [0]{}%
\providecommand \bibitemNoStop [0]{.\EOS\space}%
\providecommand \EOS [0]{\spacefactor3000\relax}%
\providecommand \BibitemShut  [1]{\csname bibitem#1\endcsname}%
\let\auto@bib@innerbib\@empty
\bibitem [{\citenamefont {Titulaer}\ and\ \citenamefont
  {Glauber}(1965)}]{titulaer65}%
  \BibitemOpen
  \bibfield  {author} {\bibinfo {author} {\bibfnamefont {U.~M.}\ \bibnamefont
  {Titulaer}}\ and\ \bibinfo {author} {\bibfnamefont {R.~J.}\ \bibnamefont
  {Glauber}},\ }\bibfield  {title} {\bibinfo {title} {Correlation functions for
  coherent fields},\ }\href {https://doi.org/10.1103/PhysRev.140.B676}
  {\bibfield  {journal} {\bibinfo  {journal} {Phys. Rev.}\ }\textbf {\bibinfo
  {volume} {140}},\ \bibinfo {pages} {B676} (\bibinfo {year}
  {1965})}\BibitemShut {NoStop}%
\bibitem [{\citenamefont {Mandel}(1986)}]{mandel86}%
  \BibitemOpen
  \bibfield  {author} {\bibinfo {author} {\bibfnamefont {L.}~\bibnamefont
  {Mandel}},\ }\bibfield  {title} {\bibinfo {title} {Non-classical states of
  the electromagnetic field},\ }\href
  {https://doi.org/10.1088/0031-8949/1986/t12/005} {\bibfield  {journal}
  {\bibinfo  {journal} {Phys. Scr.}\ }\textbf {\bibinfo {volume} {T12}},\
  \bibinfo {pages} {34} (\bibinfo {year} {1986})}\BibitemShut {NoStop}%
\bibitem [{\citenamefont {Mandel}\ and\ \citenamefont
  {Wolf}(1995)}]{mandel_book}%
  \BibitemOpen
  \bibfield  {author} {\bibinfo {author} {\bibfnamefont {L.}~\bibnamefont
  {Mandel}}\ and\ \bibinfo {author} {\bibfnamefont {E.}~\bibnamefont {Wolf}},\
  }\href@noop {} {\emph {\bibinfo {title} {Optical Coherence and Quantum
  Optics}}}\ (\bibinfo  {publisher} {Cambridge University Press, Cambridge},\
  \bibinfo {year} {1995})\BibitemShut {NoStop}%
\bibitem [{\citenamefont {Vogel}\ and\ \citenamefont
  {Welsch}(2006)}]{vogel_book}%
  \BibitemOpen
  \bibfield  {author} {\bibinfo {author} {\bibfnamefont {W.}~\bibnamefont
  {Vogel}}\ and\ \bibinfo {author} {\bibfnamefont {D.-G.}\ \bibnamefont
  {Welsch}},\ }\href@noop {} {\emph {\bibinfo {title} {Quantum Optics}}}\
  (\bibinfo  {publisher} {Wiley-VCH},\ \bibinfo {address} {Weinheim},\ \bibinfo
  {year} {2006})\BibitemShut {NoStop}%
\bibitem [{\citenamefont {Agarwal}(2013)}]{agarwal_book}%
  \BibitemOpen
  \bibfield  {author} {\bibinfo {author} {\bibfnamefont {G.~S.}\ \bibnamefont
  {Agarwal}},\ }\href@noop {} {\emph {\bibinfo {title} {Quantum Optics}}}\
  (\bibinfo  {publisher} {Cambridge University Press, Cambridge},\ \bibinfo
  {year} {2013})\BibitemShut {NoStop}%
\bibitem [{\citenamefont {Schnabel}(2017)}]{Schnabel2017}%
  \BibitemOpen
  \bibfield  {author} {\bibinfo {author} {\bibfnamefont {R.}~\bibnamefont
  {Schnabel}},\ }\bibfield  {title} {\bibinfo {title} {Squeezed states of light
  and their applications in laser interferometers},\ }\href
  {https://doi.org/https://doi.org/10.1016/j.physrep.2017.04.001} {\bibfield
  {journal} {\bibinfo  {journal} {Phys. Rep.}\ }\textbf {\bibinfo {volume}
  {684}},\ \bibinfo {pages} {1} (\bibinfo {year} {2017})}\BibitemShut {NoStop}%
\bibitem [{\citenamefont {Sperling}\ and\ \citenamefont
  {Walmsley}(2018{\natexlab{a}})}]{sperling2018a}%
  \BibitemOpen
  \bibfield  {author} {\bibinfo {author} {\bibfnamefont {J.}~\bibnamefont
  {Sperling}}\ and\ \bibinfo {author} {\bibfnamefont {I.~A.}\ \bibnamefont
  {Walmsley}},\ }\bibfield  {title} {\bibinfo {title} {Quasiprobability
  representation of quantum coherence},\ }\href
  {https://doi.org/10.1103/PhysRevA.97.062327} {\bibfield  {journal} {\bibinfo
  {journal} {Phys. Rev. A}\ }\textbf {\bibinfo {volume} {97}},\ \bibinfo
  {pages} {062327} (\bibinfo {year} {2018}{\natexlab{a}})}\BibitemShut
  {NoStop}%
\bibitem [{\citenamefont {Sperling}\ and\ \citenamefont
  {Walmsley}(2018{\natexlab{b}})}]{sperling2018b}%
  \BibitemOpen
  \bibfield  {author} {\bibinfo {author} {\bibfnamefont {J.}~\bibnamefont
  {Sperling}}\ and\ \bibinfo {author} {\bibfnamefont {I.~A.}\ \bibnamefont
  {Walmsley}},\ }\bibfield  {title} {\bibinfo {title} {Quasistates and
  quasiprobabilities},\ }\href {https://doi.org/10.1103/PhysRevA.98.042122}
  {\bibfield  {journal} {\bibinfo  {journal} {Phys. Rev. A}\ }\textbf {\bibinfo
  {volume} {98}},\ \bibinfo {pages} {042122} (\bibinfo {year}
  {2018}{\natexlab{b}})}\BibitemShut {NoStop}%
\bibitem [{\citenamefont {Sperling}\ and\ \citenamefont
  {Vogel}(2020)}]{sperling2020}%
  \BibitemOpen
  \bibfield  {author} {\bibinfo {author} {\bibfnamefont {J.}~\bibnamefont
  {Sperling}}\ and\ \bibinfo {author} {\bibfnamefont {W.}~\bibnamefont
  {Vogel}},\ }\bibfield  {title} {\bibinfo {title} {Quasiprobability
  distributions for quantum-optical coherence and beyond},\ }\href
  {https://doi.org/10.1088/1402-4896/ab5501} {\bibfield  {journal} {\bibinfo
  {journal} {Phys. Scr.}\ }\textbf {\bibinfo {volume} {95}},\ \bibinfo {pages}
  {034007} (\bibinfo {year} {2020})}\BibitemShut {NoStop}%
\bibitem [{\citenamefont {Glauber}(1963)}]{glauber63c}%
  \BibitemOpen
  \bibfield  {author} {\bibinfo {author} {\bibfnamefont {R.~J.}\ \bibnamefont
  {Glauber}},\ }\bibfield  {title} {\bibinfo {title} {Coherent and incoherent
  states of the radiation field},\ }\href
  {https://doi.org/10.1103/PhysRev.131.2766} {\bibfield  {journal} {\bibinfo
  {journal} {Phys. Rev.}\ }\textbf {\bibinfo {volume} {131}},\ \bibinfo {pages}
  {2766} (\bibinfo {year} {1963})}\BibitemShut {NoStop}%
\bibitem [{\citenamefont {Sudarshan}(1963)}]{sudarshan63}%
  \BibitemOpen
  \bibfield  {author} {\bibinfo {author} {\bibfnamefont {E.~C.~G.}\
  \bibnamefont {Sudarshan}},\ }\bibfield  {title} {\bibinfo {title}
  {Equivalence of semiclassical and quantum mechanical descriptions of
  statistical light beams},\ }\href
  {https://doi.org/10.1103/PhysRevLett.10.277} {\bibfield  {journal} {\bibinfo
  {journal} {Phys. Rev. Lett.}\ }\textbf {\bibinfo {volume} {10}},\ \bibinfo
  {pages} {277} (\bibinfo {year} {1963})}\BibitemShut {NoStop}%
\bibitem [{\citenamefont {Kiesel}\ \emph {et~al.}(2008)\citenamefont {Kiesel},
  \citenamefont {Vogel}, \citenamefont {Parigi}, \citenamefont {Zavatta},\ and\
  \citenamefont {Bellini}}]{kiesel08}%
  \BibitemOpen
  \bibfield  {author} {\bibinfo {author} {\bibfnamefont {T.}~\bibnamefont
  {Kiesel}}, \bibinfo {author} {\bibfnamefont {W.}~\bibnamefont {Vogel}},
  \bibinfo {author} {\bibfnamefont {V.}~\bibnamefont {Parigi}}, \bibinfo
  {author} {\bibfnamefont {A.}~\bibnamefont {Zavatta}},\ and\ \bibinfo {author}
  {\bibfnamefont {M.}~\bibnamefont {Bellini}},\ }\bibfield  {title} {\bibinfo
  {title} {Experimental determination of a nonclassical {Glauber-Sudarshan $P$}
  function},\ }\href {https://doi.org/10.1103/PhysRevA.78.021804} {\bibfield
  {journal} {\bibinfo  {journal} {Phys. Rev. A}\ }\textbf {\bibinfo {volume}
  {78}},\ \bibinfo {pages} {021804(R)} (\bibinfo {year} {2008})}\BibitemShut
  {NoStop}%
\bibitem [{\citenamefont {Kiesel}\ \emph {et~al.}(2011)\citenamefont {Kiesel},
  \citenamefont {Vogel}, \citenamefont {Bellini},\ and\ \citenamefont
  {Zavatta}}]{kiesel11b}%
  \BibitemOpen
  \bibfield  {author} {\bibinfo {author} {\bibfnamefont {T.}~\bibnamefont
  {Kiesel}}, \bibinfo {author} {\bibfnamefont {W.}~\bibnamefont {Vogel}},
  \bibinfo {author} {\bibfnamefont {M.}~\bibnamefont {Bellini}},\ and\ \bibinfo
  {author} {\bibfnamefont {A.}~\bibnamefont {Zavatta}},\ }\bibfield  {title}
  {\bibinfo {title} {Nonclassicality quasiprobability of single-photon-added
  thermal states},\ }\href {https://doi.org/10.1103/PhysRevA.83.032116}
  {\bibfield  {journal} {\bibinfo  {journal} {Phys. Rev. A}\ }\textbf {\bibinfo
  {volume} {83}},\ \bibinfo {pages} {032116} (\bibinfo {year}
  {2011})}\BibitemShut {NoStop}%
\bibitem [{\citenamefont {Kiesel}\ and\ \citenamefont
  {Vogel}(2010)}]{kiesel10}%
  \BibitemOpen
  \bibfield  {author} {\bibinfo {author} {\bibfnamefont {T.}~\bibnamefont
  {Kiesel}}\ and\ \bibinfo {author} {\bibfnamefont {W.}~\bibnamefont {Vogel}},\
  }\bibfield  {title} {\bibinfo {title} {Nonclassicality filters and
  quasiprobabilities},\ }\href {https://doi.org/10.1103/PhysRevA.82.032107}
  {\bibfield  {journal} {\bibinfo  {journal} {Phys. Rev. A}\ }\textbf {\bibinfo
  {volume} {82}},\ \bibinfo {pages} {032107} (\bibinfo {year}
  {2010})}\BibitemShut {NoStop}%
\bibitem [{\citenamefont {Agudelo}\ \emph {et~al.}(2013)\citenamefont
  {Agudelo}, \citenamefont {Sperling},\ and\ \citenamefont
  {Vogel}}]{Agudelo2013}%
  \BibitemOpen
  \bibfield  {author} {\bibinfo {author} {\bibfnamefont {E.}~\bibnamefont
  {Agudelo}}, \bibinfo {author} {\bibfnamefont {J.}~\bibnamefont {Sperling}},\
  and\ \bibinfo {author} {\bibfnamefont {W.}~\bibnamefont {Vogel}},\ }\bibfield
   {title} {\bibinfo {title} {Quasiprobabilities for multipartite quantum
  correlations of light},\ }\href {https://doi.org/10.1103/PhysRevA.87.033811}
  {\bibfield  {journal} {\bibinfo  {journal} {Phys. Rev. A}\ }\textbf {\bibinfo
  {volume} {87}},\ \bibinfo {pages} {033811} (\bibinfo {year}
  {2013})}\BibitemShut {NoStop}%
\bibitem [{\citenamefont {Luis}\ \emph {et~al.}(2015)\citenamefont {Luis},
  \citenamefont {Sperling},\ and\ \citenamefont {Vogel}}]{luis15}%
  \BibitemOpen
  \bibfield  {author} {\bibinfo {author} {\bibfnamefont {A.}~\bibnamefont
  {Luis}}, \bibinfo {author} {\bibfnamefont {J.}~\bibnamefont {Sperling}},\
  and\ \bibinfo {author} {\bibfnamefont {W.}~\bibnamefont {Vogel}},\ }\bibfield
   {title} {\bibinfo {title} {Nonclassicality phase-space functions: More
  insight with fewer detectors},\ }\href
  {https://doi.org/10.1103/PhysRevLett.114.103602} {\bibfield  {journal}
  {\bibinfo  {journal} {Phys. Rev. Lett.}\ }\textbf {\bibinfo {volume} {114}},\
  \bibinfo {pages} {103602} (\bibinfo {year} {2015})}\BibitemShut {NoStop}%
\bibitem [{\citenamefont {Bohmann}\ and\ \citenamefont
  {Agudelo}(2020)}]{bohmann2020}%
  \BibitemOpen
  \bibfield  {author} {\bibinfo {author} {\bibfnamefont {M.}~\bibnamefont
  {Bohmann}}\ and\ \bibinfo {author} {\bibfnamefont {E.}~\bibnamefont
  {Agudelo}},\ }\bibfield  {title} {\bibinfo {title} {Phase-space inequalities
  beyond negativities},\ }\href
  {https://doi.org/10.1103/PhysRevLett.124.133601} {\bibfield  {journal}
  {\bibinfo  {journal} {Phys. Rev. Lett.}\ }\textbf {\bibinfo {volume} {124}},\
  \bibinfo {pages} {133601} (\bibinfo {year} {2020})}\BibitemShut {NoStop}%
\bibitem [{\citenamefont {Bohmann}\ \emph {et~al.}(2020)\citenamefont
  {Bohmann}, \citenamefont {Agudelo},\ and\ \citenamefont
  {Sperling}}]{bohmann2020b}%
  \BibitemOpen
  \bibfield  {author} {\bibinfo {author} {\bibfnamefont {M.}~\bibnamefont
  {Bohmann}}, \bibinfo {author} {\bibfnamefont {E.}~\bibnamefont {Agudelo}},\
  and\ \bibinfo {author} {\bibfnamefont {J.}~\bibnamefont {Sperling}},\
  }\bibfield  {title} {\bibinfo {title} {Probing nonclassicality with matrices
  of phase-space distributions},\ }\href
  {https://doi.org/10.22331/q-2020-10-15-343} {\bibfield  {journal} {\bibinfo
  {journal} {{Quantum}}\ }\textbf {\bibinfo {volume} {4}},\ \bibinfo {pages}
  {343} (\bibinfo {year} {2020})}\BibitemShut {NoStop}%
\bibitem [{\citenamefont {Hillery}(1987)}]{Hillery1987}%
  \BibitemOpen
  \bibfield  {author} {\bibinfo {author} {\bibfnamefont {M.}~\bibnamefont
  {Hillery}},\ }\bibfield  {title} {\bibinfo {title} {Nonclassical distance in
  quantum optics},\ }\href {https://doi.org/10.1103/PhysRevA.35.725} {\bibfield
   {journal} {\bibinfo  {journal} {Phys. Rev. A}\ }\textbf {\bibinfo {volume}
  {35}},\ \bibinfo {pages} {725} (\bibinfo {year} {1987})}\BibitemShut
  {NoStop}%
\bibitem [{\citenamefont {Lee}(1991)}]{Lee1991}%
  \BibitemOpen
  \bibfield  {author} {\bibinfo {author} {\bibfnamefont {C.~T.}\ \bibnamefont
  {Lee}},\ }\bibfield  {title} {\bibinfo {title} {Measure of the
  nonclassicality of nonclassical states},\ }\href
  {https://doi.org/10.1103/PhysRevA.44.R2775} {\bibfield  {journal} {\bibinfo
  {journal} {Phys. Rev. A}\ }\textbf {\bibinfo {volume} {44}},\ \bibinfo
  {pages} {R2775} (\bibinfo {year} {1991})}\BibitemShut {NoStop}%
\bibitem [{\citenamefont {Asb\'oth}\ \emph {et~al.}(2005)\citenamefont
  {Asb\'oth}, \citenamefont {Calsamiglia},\ and\ \citenamefont
  {Ritsch}}]{Asboth2005}%
  \BibitemOpen
  \bibfield  {author} {\bibinfo {author} {\bibfnamefont {J.~K.}\ \bibnamefont
  {Asb\'oth}}, \bibinfo {author} {\bibfnamefont {J.}~\bibnamefont
  {Calsamiglia}},\ and\ \bibinfo {author} {\bibfnamefont {H.}~\bibnamefont
  {Ritsch}},\ }\bibfield  {title} {\bibinfo {title} {Computable measure of
  nonclassicality for light},\ }\href
  {https://doi.org/10.1103/PhysRevLett.94.173602} {\bibfield  {journal}
  {\bibinfo  {journal} {Phys. Rev. Lett.}\ }\textbf {\bibinfo {volume} {94}},\
  \bibinfo {pages} {173602} (\bibinfo {year} {2005})}\BibitemShut {NoStop}%
\bibitem [{\citenamefont {Miranowicz}\ \emph
  {et~al.}(2015{\natexlab{a}})\citenamefont {Miranowicz}, \citenamefont
  {Bartkiewicz}, \citenamefont {Pathak}, \citenamefont
  {Pe\ifmmode~\check{r}\else \v{r}\fi{}ina}, \citenamefont {Chen},\ and\
  \citenamefont {Nori}}]{Miranowicz2015a}%
  \BibitemOpen
  \bibfield  {author} {\bibinfo {author} {\bibfnamefont {A.}~\bibnamefont
  {Miranowicz}}, \bibinfo {author} {\bibfnamefont {K.}~\bibnamefont
  {Bartkiewicz}}, \bibinfo {author} {\bibfnamefont {A.}~\bibnamefont {Pathak}},
  \bibinfo {author} {\bibfnamefont {J.}~\bibnamefont {Pe\ifmmode~\check{r}\else
  \v{r}\fi{}ina}}, \bibinfo {author} {\bibfnamefont {Y.-N.}\ \bibnamefont
  {Chen}},\ and\ \bibinfo {author} {\bibfnamefont {F.}~\bibnamefont {Nori}},\
  }\bibfield  {title} {\bibinfo {title} {Statistical mixtures of states can be
  more quantum than their superpositions: Comparison of nonclassicality
  measures for single-qubit states},\ }\href
  {https://doi.org/10.1103/PhysRevA.91.042309} {\bibfield  {journal} {\bibinfo
  {journal} {Phys. Rev. A}\ }\textbf {\bibinfo {volume} {91}},\ \bibinfo
  {pages} {042309} (\bibinfo {year} {2015}{\natexlab{a}})}\BibitemShut
  {NoStop}%
\bibitem [{\citenamefont {Miranowicz}\ \emph
  {et~al.}(2015{\natexlab{b}})\citenamefont {Miranowicz}, \citenamefont
  {Bartkiewicz}, \citenamefont {Lambert}, \citenamefont {Chen},\ and\
  \citenamefont {Nori}}]{Miranowicz2015b}%
  \BibitemOpen
  \bibfield  {author} {\bibinfo {author} {\bibfnamefont {A.}~\bibnamefont
  {Miranowicz}}, \bibinfo {author} {\bibfnamefont {K.}~\bibnamefont
  {Bartkiewicz}}, \bibinfo {author} {\bibfnamefont {N.}~\bibnamefont
  {Lambert}}, \bibinfo {author} {\bibfnamefont {Y.-N.}\ \bibnamefont {Chen}},\
  and\ \bibinfo {author} {\bibfnamefont {F.}~\bibnamefont {Nori}},\ }\bibfield
  {title} {\bibinfo {title} {Increasing relative nonclassicality quantified by
  standard entanglement potentials by dissipation and unbalanced beam
  splitting},\ }\href {https://doi.org/10.1103/PhysRevA.92.062314} {\bibfield
  {journal} {\bibinfo  {journal} {Phys. Rev. A}\ }\textbf {\bibinfo {volume}
  {92}},\ \bibinfo {pages} {062314} (\bibinfo {year}
  {2015}{\natexlab{b}})}\BibitemShut {NoStop}%
\bibitem [{\citenamefont {Yadin}\ \emph {et~al.}(2018)\citenamefont {Yadin},
  \citenamefont {Binder}, \citenamefont {Thompson}, \citenamefont
  {Narasimhachar}, \citenamefont {Gu},\ and\ \citenamefont {Kim}}]{Yadin2019}%
  \BibitemOpen
  \bibfield  {author} {\bibinfo {author} {\bibfnamefont {B.}~\bibnamefont
  {Yadin}}, \bibinfo {author} {\bibfnamefont {F.~C.}\ \bibnamefont {Binder}},
  \bibinfo {author} {\bibfnamefont {J.}~\bibnamefont {Thompson}}, \bibinfo
  {author} {\bibfnamefont {V.}~\bibnamefont {Narasimhachar}}, \bibinfo {author}
  {\bibfnamefont {M.}~\bibnamefont {Gu}},\ and\ \bibinfo {author}
  {\bibfnamefont {M.~S.}\ \bibnamefont {Kim}},\ }\bibfield  {title} {\bibinfo
  {title} {Operational resource theory of continuous-variable
  nonclassicality},\ }\href {https://doi.org/10.1103/PhysRevX.8.041038}
  {\bibfield  {journal} {\bibinfo  {journal} {Phys. Rev. X}\ }\textbf {\bibinfo
  {volume} {8}},\ \bibinfo {pages} {041038} (\bibinfo {year}
  {2018})}\BibitemShut {NoStop}%
\bibitem [{\citenamefont {Luo}\ and\ \citenamefont {Zhang}(2019)}]{Luo2019}%
  \BibitemOpen
  \bibfield  {author} {\bibinfo {author} {\bibfnamefont {S.}~\bibnamefont
  {Luo}}\ and\ \bibinfo {author} {\bibfnamefont {Y.}~\bibnamefont {Zhang}},\
  }\bibfield  {title} {\bibinfo {title} {Quantifying nonclassicality via
  {W}igner-{Y}anase skew information},\ }\href
  {https://doi.org/10.1103/PhysRevA.100.032116} {\bibfield  {journal} {\bibinfo
   {journal} {Phys. Rev. A}\ }\textbf {\bibinfo {volume} {100}},\ \bibinfo
  {pages} {032116} (\bibinfo {year} {2019})}\BibitemShut {NoStop}%
\bibitem [{\citenamefont {Stoler}(1970)}]{Stoler1970}%
  \BibitemOpen
  \bibfield  {author} {\bibinfo {author} {\bibfnamefont {D.}~\bibnamefont
  {Stoler}},\ }\bibfield  {title} {\bibinfo {title} {Equivalence classes of
  minimum uncertainty packets},\ }\href
  {https://doi.org/10.1103/PhysRevD.1.3217} {\bibfield  {journal} {\bibinfo
  {journal} {Phys. Rev. D}\ }\textbf {\bibinfo {volume} {1}},\ \bibinfo {pages}
  {3217} (\bibinfo {year} {1970})}\BibitemShut {NoStop}%
\bibitem [{\citenamefont {Stoler}(1971)}]{Stoler1971}%
  \BibitemOpen
  \bibfield  {author} {\bibinfo {author} {\bibfnamefont {D.}~\bibnamefont
  {Stoler}},\ }\bibfield  {title} {\bibinfo {title} {Equivalence classes of
  minimum-uncertainty packets. ii},\ }\href
  {https://doi.org/10.1103/PhysRevD.4.1925} {\bibfield  {journal} {\bibinfo
  {journal} {Phys. Rev. D}\ }\textbf {\bibinfo {volume} {4}},\ \bibinfo {pages}
  {1925} (\bibinfo {year} {1971})}\BibitemShut {NoStop}%
\bibitem [{\citenamefont {Mandel}(1979)}]{mandel79}%
  \BibitemOpen
  \bibfield  {author} {\bibinfo {author} {\bibfnamefont {L.}~\bibnamefont
  {Mandel}},\ }\bibfield  {title} {\bibinfo {title} {Sub-{P}oissonian photon
  statistics in resonance fluorescence},\ }\href
  {https://doi.org/10.1364/OL.4.000205} {\bibfield  {journal} {\bibinfo
  {journal} {Opt. Lett.}\ }\textbf {\bibinfo {volume} {4}},\ \bibinfo {pages}
  {205} (\bibinfo {year} {1979})}\BibitemShut {NoStop}%
\bibitem [{\citenamefont {Reid}\ and\ \citenamefont {Walls}(1986)}]{reid1986}%
  \BibitemOpen
  \bibfield  {author} {\bibinfo {author} {\bibfnamefont {M.~D.}\ \bibnamefont
  {Reid}}\ and\ \bibinfo {author} {\bibfnamefont {D.~F.}\ \bibnamefont
  {Walls}},\ }\bibfield  {title} {\bibinfo {title} {Violations of classical
  inequalities in quantum optics},\ }\href
  {https://doi.org/10.1103/PhysRevA.34.1260} {\bibfield  {journal} {\bibinfo
  {journal} {Phys. Rev. A}\ }\textbf {\bibinfo {volume} {34}},\ \bibinfo
  {pages} {1260} (\bibinfo {year} {1986})}\BibitemShut {NoStop}%
\bibitem [{\citenamefont {Wu}\ \emph {et~al.}(1986)\citenamefont {Wu},
  \citenamefont {Kimble}, \citenamefont {Hall},\ and\ \citenamefont
  {Wu}}]{Wu1986}%
  \BibitemOpen
  \bibfield  {author} {\bibinfo {author} {\bibfnamefont {L.-A.}\ \bibnamefont
  {Wu}}, \bibinfo {author} {\bibfnamefont {H.~J.}\ \bibnamefont {Kimble}},
  \bibinfo {author} {\bibfnamefont {J.~L.}\ \bibnamefont {Hall}},\ and\
  \bibinfo {author} {\bibfnamefont {H.}~\bibnamefont {Wu}},\ }\bibfield
  {title} {\bibinfo {title} {Generation of squeezed states by parametric down
  conversion},\ }\href {https://doi.org/10.1103/PhysRevLett.57.2520} {\bibfield
   {journal} {\bibinfo  {journal} {Phys. Rev. Lett.}\ }\textbf {\bibinfo
  {volume} {57}},\ \bibinfo {pages} {2520} (\bibinfo {year}
  {1986})}\BibitemShut {NoStop}%
\bibitem [{\citenamefont {Wu}\ \emph {et~al.}(1987)\citenamefont {Wu},
  \citenamefont {Xiao},\ and\ \citenamefont {Kimble}}]{Wu1987}%
  \BibitemOpen
  \bibfield  {author} {\bibinfo {author} {\bibfnamefont {L.-A.}\ \bibnamefont
  {Wu}}, \bibinfo {author} {\bibfnamefont {M.}~\bibnamefont {Xiao}},\ and\
  \bibinfo {author} {\bibfnamefont {H.~J.}\ \bibnamefont {Kimble}},\ }\bibfield
   {title} {\bibinfo {title} {Squeezed states of light from an optical
  parametric oscillator},\ }\href {https://doi.org/10.1364/JOSAB.4.001465}
  {\bibfield  {journal} {\bibinfo  {journal} {J. Opt. Soc. Am. B}\ }\textbf
  {\bibinfo {volume} {4}},\ \bibinfo {pages} {1465} (\bibinfo {year}
  {1987})}\BibitemShut {NoStop}%
\bibitem [{\citenamefont {Vahlbruch}\ \emph {et~al.}(2016)\citenamefont
  {Vahlbruch}, \citenamefont {Mehmet}, \citenamefont {Danzmann},\ and\
  \citenamefont {Schnabel}}]{Vahlbruch}%
  \BibitemOpen
  \bibfield  {author} {\bibinfo {author} {\bibfnamefont {H.}~\bibnamefont
  {Vahlbruch}}, \bibinfo {author} {\bibfnamefont {M.}~\bibnamefont {Mehmet}},
  \bibinfo {author} {\bibfnamefont {K.}~\bibnamefont {Danzmann}},\ and\
  \bibinfo {author} {\bibfnamefont {R.}~\bibnamefont {Schnabel}},\ }\bibfield
  {title} {\bibinfo {title} {Detection of 15 {dB} squeezed states of light and
  their application for the absolute calibration of photoelectric quantum
  efficiency},\ }\href {https://doi.org/10.1103/PhysRevLett.117.110801}
  {\bibfield  {journal} {\bibinfo  {journal} {Phys. Rev. Lett.}\ }\textbf
  {\bibinfo {volume} {117}},\ \bibinfo {pages} {110801} (\bibinfo {year}
  {2016})}\BibitemShut {NoStop}%
\bibitem [{\citenamefont {Agarwal}\ and\ \citenamefont
  {Tara}(1992)}]{agarwal92}%
  \BibitemOpen
  \bibfield  {author} {\bibinfo {author} {\bibfnamefont {G.~S.}\ \bibnamefont
  {Agarwal}}\ and\ \bibinfo {author} {\bibfnamefont {K.}~\bibnamefont {Tara}},\
  }\bibfield  {title} {\bibinfo {title} {Nonclassical character of states
  exhibiting no squeezing or sub-{P}oissonian statistics},\ }\href
  {https://doi.org/10.1103/PhysRevA.46.485} {\bibfield  {journal} {\bibinfo
  {journal} {Phys. Rev. A}\ }\textbf {\bibinfo {volume} {46}},\ \bibinfo
  {pages} {485} (\bibinfo {year} {1992})}\BibitemShut {NoStop}%
\bibitem [{\citenamefont {Agarwal}(1993)}]{agarwal93}%
  \BibitemOpen
  \bibfield  {author} {\bibinfo {author} {\bibfnamefont {G.}~\bibnamefont
  {Agarwal}},\ }\bibfield  {title} {\bibinfo {title} {Nonclassical
  characteristics of the marginals for the radiation field},\ }\href
  {https://doi.org/http://dx.doi.org/10.1016/0030-4018(93)90059-E} {\bibfield
  {journal} {\bibinfo  {journal} {Opt. Commun.}\ }\textbf {\bibinfo {volume}
  {95}},\ \bibinfo {pages} {109 } (\bibinfo {year} {1993})}\BibitemShut
  {NoStop}%
\bibitem [{\citenamefont {Klyshko}(1996)}]{klyshko1996}%
  \BibitemOpen
  \bibfield  {author} {\bibinfo {author} {\bibfnamefont {D.~N.}\ \bibnamefont
  {Klyshko}},\ }\bibfield  {title} {\bibinfo {title} {Observable signs of
  nonclassical light},\ }\href
  {https://doi.org/https://doi.org/10.1016/0375-9601(96)00091-6} {\bibfield
  {journal} {\bibinfo  {journal} {Phys. Lett. A}\ }\textbf {\bibinfo {volume}
  {213}},\ \bibinfo {pages} {7} (\bibinfo {year} {1996})}\BibitemShut {NoStop}%
\bibitem [{\citenamefont {Vogel}(2000)}]{vogel00}%
  \BibitemOpen
  \bibfield  {author} {\bibinfo {author} {\bibfnamefont {W.}~\bibnamefont
  {Vogel}},\ }\bibfield  {title} {\bibinfo {title} {Nonclassical states: An
  observable criterion},\ }\href {https://doi.org/10.1103/PhysRevLett.84.1849}
  {\bibfield  {journal} {\bibinfo  {journal} {Phys. Rev. Lett.}\ }\textbf
  {\bibinfo {volume} {84}},\ \bibinfo {pages} {1849} (\bibinfo {year}
  {2000})}\BibitemShut {NoStop}%
\bibitem [{\citenamefont {Richter}\ and\ \citenamefont
  {Vogel}(2002)}]{richter02}%
  \BibitemOpen
  \bibfield  {author} {\bibinfo {author} {\bibfnamefont {T.}~\bibnamefont
  {Richter}}\ and\ \bibinfo {author} {\bibfnamefont {W.}~\bibnamefont
  {Vogel}},\ }\bibfield  {title} {\bibinfo {title} {Nonclassicality of quantum
  states: A hierarchy of observable conditions},\ }\href
  {https://doi.org/10.1103/PhysRevLett.89.283601} {\bibfield  {journal}
  {\bibinfo  {journal} {Phys. Rev. Lett.}\ }\textbf {\bibinfo {volume} {89}},\
  \bibinfo {pages} {283601} (\bibinfo {year} {2002})}\BibitemShut {NoStop}%
\bibitem [{\citenamefont {Rivas}\ and\ \citenamefont {Luis}(2009)}]{rivas2009}%
  \BibitemOpen
  \bibfield  {author} {\bibinfo {author} {\bibfnamefont {A.}~\bibnamefont
  {Rivas}}\ and\ \bibinfo {author} {\bibfnamefont {A.}~\bibnamefont {Luis}},\
  }\bibfield  {title} {\bibinfo {title} {Nonclassicality of states and
  measurements by breaking classical bounds on statistics},\ }\href
  {https://doi.org/10.1103/PhysRevA.79.042105} {\bibfield  {journal} {\bibinfo
  {journal} {Phys. Rev. A}\ }\textbf {\bibinfo {volume} {79}},\ \bibinfo
  {pages} {042105} (\bibinfo {year} {2009})}\BibitemShut {NoStop}%
\bibitem [{\citenamefont {Sperling}\ \emph
  {et~al.}(2012{\natexlab{a}})\citenamefont {Sperling}, \citenamefont {Vogel},\
  and\ \citenamefont {Agarwal}}]{sperling12c}%
  \BibitemOpen
  \bibfield  {author} {\bibinfo {author} {\bibfnamefont {J.}~\bibnamefont
  {Sperling}}, \bibinfo {author} {\bibfnamefont {W.}~\bibnamefont {Vogel}},\
  and\ \bibinfo {author} {\bibfnamefont {G.~S.}\ \bibnamefont {Agarwal}},\
  }\bibfield  {title} {\bibinfo {title} {Sub-binomial light},\ }\href
  {https://doi.org/10.1103/PhysRevLett.109.093601} {\bibfield  {journal}
  {\bibinfo  {journal} {Phys. Rev. Lett.}\ }\textbf {\bibinfo {volume} {109}},\
  \bibinfo {pages} {093601} (\bibinfo {year} {2012}{\natexlab{a}})}\BibitemShut
  {NoStop}%
\bibitem [{\citenamefont {Bartley}\ \emph {et~al.}(2013)\citenamefont
  {Bartley}, \citenamefont {Donati}, \citenamefont {Jin}, \citenamefont
  {Datta}, \citenamefont {Barbieri},\ and\ \citenamefont
  {Walmsley}}]{bartley13}%
  \BibitemOpen
  \bibfield  {author} {\bibinfo {author} {\bibfnamefont {T.~J.}\ \bibnamefont
  {Bartley}}, \bibinfo {author} {\bibfnamefont {G.}~\bibnamefont {Donati}},
  \bibinfo {author} {\bibfnamefont {X.-M.}\ \bibnamefont {Jin}}, \bibinfo
  {author} {\bibfnamefont {A.}~\bibnamefont {Datta}}, \bibinfo {author}
  {\bibfnamefont {M.}~\bibnamefont {Barbieri}},\ and\ \bibinfo {author}
  {\bibfnamefont {I.~A.}\ \bibnamefont {Walmsley}},\ }\bibfield  {title}
  {\bibinfo {title} {Direct observation of sub-binomial light},\ }\href
  {https://doi.org/10.1103/PhysRevLett.110.173602} {\bibfield  {journal}
  {\bibinfo  {journal} {Phys. Rev. Lett.}\ }\textbf {\bibinfo {volume} {110}},\
  \bibinfo {pages} {173602} (\bibinfo {year} {2013})}\BibitemShut {NoStop}%
\bibitem [{\citenamefont {Sperling}\ \emph {et~al.}(2013)\citenamefont
  {Sperling}, \citenamefont {Vogel},\ and\ \citenamefont
  {Agarwal}}]{sperling13b}%
  \BibitemOpen
  \bibfield  {author} {\bibinfo {author} {\bibfnamefont {J.}~\bibnamefont
  {Sperling}}, \bibinfo {author} {\bibfnamefont {W.}~\bibnamefont {Vogel}},\
  and\ \bibinfo {author} {\bibfnamefont {G.~S.}\ \bibnamefont {Agarwal}},\
  }\bibfield  {title} {\bibinfo {title} {Correlation measurements with on-off
  detectors},\ }\href {https://doi.org/10.1103/PhysRevA.88.043821} {\bibfield
  {journal} {\bibinfo  {journal} {Phys. Rev. A}\ }\textbf {\bibinfo {volume}
  {88}},\ \bibinfo {pages} {043821} (\bibinfo {year} {2013})}\BibitemShut
  {NoStop}%
\bibitem [{\citenamefont {Park}\ \emph {et~al.}(2015)\citenamefont {Park},
  \citenamefont {Zhang}, \citenamefont {Lee}, \citenamefont {Ji}, \citenamefont
  {Um}, \citenamefont {Lv}, \citenamefont {Kim},\ and\ \citenamefont
  {Nha}}]{park2015a}%
  \BibitemOpen
  \bibfield  {author} {\bibinfo {author} {\bibfnamefont {J.}~\bibnamefont
  {Park}}, \bibinfo {author} {\bibfnamefont {J.}~\bibnamefont {Zhang}},
  \bibinfo {author} {\bibfnamefont {J.}~\bibnamefont {Lee}}, \bibinfo {author}
  {\bibfnamefont {S.-W.}\ \bibnamefont {Ji}}, \bibinfo {author} {\bibfnamefont
  {M.}~\bibnamefont {Um}}, \bibinfo {author} {\bibfnamefont {D.}~\bibnamefont
  {Lv}}, \bibinfo {author} {\bibfnamefont {K.}~\bibnamefont {Kim}},\ and\
  \bibinfo {author} {\bibfnamefont {H.}~\bibnamefont {Nha}},\ }\bibfield
  {title} {\bibinfo {title} {Testing nonclassicality and non-{G}aussianity in
  phase space},\ }\href {https://doi.org/10.1103/PhysRevLett.114.190402}
  {\bibfield  {journal} {\bibinfo  {journal} {Phys. Rev. Lett.}\ }\textbf
  {\bibinfo {volume} {114}},\ \bibinfo {pages} {190402} (\bibinfo {year}
  {2015})}\BibitemShut {NoStop}%
\bibitem [{\citenamefont {Park}\ and\ \citenamefont {Nha}(2015)}]{park2015b}%
  \BibitemOpen
  \bibfield  {author} {\bibinfo {author} {\bibfnamefont {J.}~\bibnamefont
  {Park}}\ and\ \bibinfo {author} {\bibfnamefont {H.}~\bibnamefont {Nha}},\
  }\bibfield  {title} {\bibinfo {title} {Demonstrating nonclassicality and
  non-{G}aussianity of single-mode fields: {Bell}-type tests using generalized
  phase-space distributions},\ }\href
  {https://doi.org/10.1103/PhysRevA.92.062134} {\bibfield  {journal} {\bibinfo
  {journal} {Phys. Rev. A}\ }\textbf {\bibinfo {volume} {92}},\ \bibinfo
  {pages} {062134} (\bibinfo {year} {2015})}\BibitemShut {NoStop}%
\bibitem [{\citenamefont {Pe\v{r}ina}\ \emph {et~al.}(2020)\citenamefont
  {Pe\v{r}ina}, \citenamefont {Mich\'{a}lek},\ and\ \citenamefont
  {Haderka}}]{Perina2020}%
  \BibitemOpen
  \bibfield  {author} {\bibinfo {author} {\bibfnamefont {J.}~\bibnamefont
  {Pe\v{r}ina}}, \bibinfo {author} {\bibfnamefont {V.}~\bibnamefont
  {Mich\'{a}lek}},\ and\ \bibinfo {author} {\bibfnamefont {O.}~\bibnamefont
  {Haderka}},\ }\bibfield  {title} {\bibinfo {title} {Non-classicality of
  optical fields as observed in photocount and photon-number distributions},\
  }\href {https://doi.org/10.1364/OE.405548} {\bibfield  {journal} {\bibinfo
  {journal} {Opt. Express}\ }\textbf {\bibinfo {volume} {28}},\ \bibinfo
  {pages} {32620} (\bibinfo {year} {2020})}\BibitemShut {NoStop}%
\bibitem [{\citenamefont {Semenov}\ and\ \citenamefont
  {Klimov}(2021)}]{Semenov2021}%
  \BibitemOpen
  \bibfield  {author} {\bibinfo {author} {\bibfnamefont {A.~A.}\ \bibnamefont
  {Semenov}}\ and\ \bibinfo {author} {\bibfnamefont {A.~B.}\ \bibnamefont
  {Klimov}},\ }\bibfield  {title} {\bibinfo {title} {Dual form of the
  phase-space classical simulation problem in quantum optics},\ }\href
  {https://doi.org/10.1088/1367-2630/ac40cc} {\bibfield  {journal} {\bibinfo
  {journal} {New J. Phys.}\ }\textbf {\bibinfo {volume} {23}},\ \bibinfo
  {pages} {123046} (\bibinfo {year} {2021})}\BibitemShut {NoStop}%
\bibitem [{\citenamefont {Innocenti}\ \emph {et~al.}(2023)\citenamefont
  {Innocenti}, \citenamefont {Lachman},\ and\ \citenamefont
  {Filip}}]{Innocenti2023}%
  \BibitemOpen
  \bibfield  {author} {\bibinfo {author} {\bibfnamefont {L.}~\bibnamefont
  {Innocenti}}, \bibinfo {author} {\bibfnamefont {L.}~\bibnamefont {Lachman}},\
  and\ \bibinfo {author} {\bibfnamefont {R.}~\bibnamefont {Filip}},\ }\bibfield
   {title} {\bibinfo {title} {Coherence-based operational nonclassicality
  criteria},\ }\href {https://doi.org/10.1103/PhysRevLett.131.160201}
  {\bibfield  {journal} {\bibinfo  {journal} {Phys. Rev. Lett.}\ }\textbf
  {\bibinfo {volume} {131}},\ \bibinfo {pages} {160201} (\bibinfo {year}
  {2023})}\BibitemShut {NoStop}%
\bibitem [{\citenamefont {Paul}\ \emph {et~al.}(1996)\citenamefont {Paul},
  \citenamefont {T\"orm\"a}, \citenamefont {Kiss},\ and\ \citenamefont
  {Jex}}]{paul1996}%
  \BibitemOpen
  \bibfield  {author} {\bibinfo {author} {\bibfnamefont {H.}~\bibnamefont
  {Paul}}, \bibinfo {author} {\bibfnamefont {P.}~\bibnamefont {T\"orm\"a}},
  \bibinfo {author} {\bibfnamefont {T.}~\bibnamefont {Kiss}},\ and\ \bibinfo
  {author} {\bibfnamefont {I.}~\bibnamefont {Jex}},\ }\bibfield  {title}
  {\bibinfo {title} {Photon chopping: New way to measure the quantum state of
  light},\ }\href {https://doi.org/10.1103/PhysRevLett.76.2464} {\bibfield
  {journal} {\bibinfo  {journal} {Phys. Rev. Lett.}\ }\textbf {\bibinfo
  {volume} {76}},\ \bibinfo {pages} {2464} (\bibinfo {year}
  {1996})}\BibitemShut {NoStop}%
\bibitem [{\citenamefont {Castelletto}\ \emph {et~al.}(2007)\citenamefont
  {Castelletto}, \citenamefont {Degiovanni}, \citenamefont {Schettini},\ and\
  \citenamefont {Migdall}}]{castelletto2007}%
  \BibitemOpen
  \bibfield  {author} {\bibinfo {author} {\bibfnamefont {S.~A.}\ \bibnamefont
  {Castelletto}}, \bibinfo {author} {\bibfnamefont {I.~P.}\ \bibnamefont
  {Degiovanni}}, \bibinfo {author} {\bibfnamefont {V.}~\bibnamefont
  {Schettini}},\ and\ \bibinfo {author} {\bibfnamefont {A.~L.}\ \bibnamefont
  {Migdall}},\ }\bibfield  {title} {\bibinfo {title} {Reduced deadtime and
  higher rate photon-counting detection using a multiplexed detector array},\
  }\href {https://doi.org/10.1080/09500340600779579} {\bibfield  {journal}
  {\bibinfo  {journal} {J. Mod. Opt.}\ }\textbf {\bibinfo {volume} {54}},\
  \bibinfo {pages} {337} (\bibinfo {year} {2007})}\BibitemShut {NoStop}%
\bibitem [{\citenamefont {Schettini}\ \emph {et~al.}(2007)\citenamefont
  {Schettini}, \citenamefont {Polyakov}, \citenamefont {Degiovanni},
  \citenamefont {Brida}, \citenamefont {Castelletto},\ and\ \citenamefont
  {Migdall}}]{schettini2007}%
  \BibitemOpen
  \bibfield  {author} {\bibinfo {author} {\bibfnamefont {V.}~\bibnamefont
  {Schettini}}, \bibinfo {author} {\bibfnamefont {S.~V.}\ \bibnamefont
  {Polyakov}}, \bibinfo {author} {\bibfnamefont {I.~P.}\ \bibnamefont
  {Degiovanni}}, \bibinfo {author} {\bibfnamefont {G.}~\bibnamefont {Brida}},
  \bibinfo {author} {\bibfnamefont {S.}~\bibnamefont {Castelletto}},\ and\
  \bibinfo {author} {\bibfnamefont {A.~L.}\ \bibnamefont {Migdall}},\
  }\bibfield  {title} {\bibinfo {title} {Implementing a multiplexed system of
  detectors for higher photon counting rates},\ }\href
  {https://doi.org/10.1109/JSTQE.2007.902846} {\bibfield  {journal} {\bibinfo
  {journal} {IEEE J. Sel. Top. Quantum Electron.}\ }\textbf {\bibinfo {volume}
  {13}},\ \bibinfo {pages} {978} (\bibinfo {year} {2007})}\BibitemShut
  {NoStop}%
\bibitem [{\citenamefont {Blanchet}\ \emph {et~al.}(2008)\citenamefont
  {Blanchet}, \citenamefont {Devaux}, \citenamefont {Furfaro},\ and\
  \citenamefont {Lantz}}]{blanchet08}%
  \BibitemOpen
  \bibfield  {author} {\bibinfo {author} {\bibfnamefont {J.-L.}\ \bibnamefont
  {Blanchet}}, \bibinfo {author} {\bibfnamefont {F.}~\bibnamefont {Devaux}},
  \bibinfo {author} {\bibfnamefont {L.}~\bibnamefont {Furfaro}},\ and\ \bibinfo
  {author} {\bibfnamefont {E.}~\bibnamefont {Lantz}},\ }\bibfield  {title}
  {\bibinfo {title} {Measurement of sub-shot-noise correlations of spatial
  fluctuations in the photon-counting regime},\ }\href
  {https://doi.org/10.1103/PhysRevLett.101.233604} {\bibfield  {journal}
  {\bibinfo  {journal} {Phys. Rev. Lett.}\ }\textbf {\bibinfo {volume} {101}},\
  \bibinfo {pages} {233604} (\bibinfo {year} {2008})}\BibitemShut {NoStop}%
\bibitem [{\citenamefont {Achilles}\ \emph {et~al.}(2003)\citenamefont
  {Achilles}, \citenamefont {Silberhorn}, \citenamefont {\'{S}liwa},
  \citenamefont {Banaszek},\ and\ \citenamefont {Walmsley}}]{achilles03}%
  \BibitemOpen
  \bibfield  {author} {\bibinfo {author} {\bibfnamefont {D.}~\bibnamefont
  {Achilles}}, \bibinfo {author} {\bibfnamefont {C.}~\bibnamefont
  {Silberhorn}}, \bibinfo {author} {\bibfnamefont {C.}~\bibnamefont
  {\'{S}liwa}}, \bibinfo {author} {\bibfnamefont {K.}~\bibnamefont
  {Banaszek}},\ and\ \bibinfo {author} {\bibfnamefont {I.~A.}\ \bibnamefont
  {Walmsley}},\ }\bibfield  {title} {\bibinfo {title} {Fiber-assisted detection
  with photon number resolution},\ }\href
  {https://doi.org/10.1364/OL.28.002387} {\bibfield  {journal} {\bibinfo
  {journal} {Opt. Lett.}\ }\textbf {\bibinfo {volume} {28}},\ \bibinfo {pages}
  {2387} (\bibinfo {year} {2003})}\BibitemShut {NoStop}%
\bibitem [{\citenamefont {Fitch}\ \emph {et~al.}(2003)\citenamefont {Fitch},
  \citenamefont {Jacobs}, \citenamefont {Pittman},\ and\ \citenamefont
  {Franson}}]{fitch03}%
  \BibitemOpen
  \bibfield  {author} {\bibinfo {author} {\bibfnamefont {M.~J.}\ \bibnamefont
  {Fitch}}, \bibinfo {author} {\bibfnamefont {B.~C.}\ \bibnamefont {Jacobs}},
  \bibinfo {author} {\bibfnamefont {T.~B.}\ \bibnamefont {Pittman}},\ and\
  \bibinfo {author} {\bibfnamefont {J.~D.}\ \bibnamefont {Franson}},\
  }\bibfield  {title} {\bibinfo {title} {Photon-number resolution using
  time-multiplexed single-photon detectors},\ }\href
  {https://doi.org/10.1103/PhysRevA.68.043814} {\bibfield  {journal} {\bibinfo
  {journal} {Phys. Rev. A}\ }\textbf {\bibinfo {volume} {68}},\ \bibinfo
  {pages} {043814} (\bibinfo {year} {2003})}\BibitemShut {NoStop}%
\bibitem [{\citenamefont {\ifmmode \check{R}\else
  \v{R}\fi{}eh\'a\ifmmode~\check{c}\else \v{c}\fi{}ek}\ \emph
  {et~al.}(2003)\citenamefont {\ifmmode \check{R}\else
  \v{R}\fi{}eh\'a\ifmmode~\check{c}\else \v{c}\fi{}ek}, \citenamefont {Hradil},
  \citenamefont {Haderka}, \citenamefont {Pe\ifmmode~\check{r}\else
  \v{r}\fi{}ina},\ and\ \citenamefont {Hamar}}]{rehacek03}%
  \BibitemOpen
  \bibfield  {author} {\bibinfo {author} {\bibfnamefont {J.}~\bibnamefont
  {\ifmmode \check{R}\else \v{R}\fi{}eh\'a\ifmmode~\check{c}\else
  \v{c}\fi{}ek}}, \bibinfo {author} {\bibfnamefont {Z.}~\bibnamefont {Hradil}},
  \bibinfo {author} {\bibfnamefont {O.}~\bibnamefont {Haderka}}, \bibinfo
  {author} {\bibfnamefont {J.}~\bibnamefont {Pe\ifmmode~\check{r}\else
  \v{r}\fi{}ina}},\ and\ \bibinfo {author} {\bibfnamefont {M.}~\bibnamefont
  {Hamar}},\ }\bibfield  {title} {\bibinfo {title} {Multiple-photon resolving
  fiber-loop detector},\ }\href {https://doi.org/10.1103/PhysRevA.67.061801}
  {\bibfield  {journal} {\bibinfo  {journal} {Phys. Rev. A}\ }\textbf {\bibinfo
  {volume} {67}},\ \bibinfo {pages} {061801(R)} (\bibinfo {year}
  {2003})}\BibitemShut {NoStop}%
\bibitem [{\citenamefont {Sperling}\ \emph
  {et~al.}(2012{\natexlab{b}})\citenamefont {Sperling}, \citenamefont {Vogel},\
  and\ \citenamefont {Agarwal}}]{sperling12a}%
  \BibitemOpen
  \bibfield  {author} {\bibinfo {author} {\bibfnamefont {J.}~\bibnamefont
  {Sperling}}, \bibinfo {author} {\bibfnamefont {W.}~\bibnamefont {Vogel}},\
  and\ \bibinfo {author} {\bibfnamefont {G.~S.}\ \bibnamefont {Agarwal}},\
  }\bibfield  {title} {\bibinfo {title} {True photocounting statistics of
  multiple on-off detectors},\ }\href
  {https://doi.org/10.1103/PhysRevA.85.023820} {\bibfield  {journal} {\bibinfo
  {journal} {Phys. Rev. A}\ }\textbf {\bibinfo {volume} {85}},\ \bibinfo
  {pages} {023820} (\bibinfo {year} {2012}{\natexlab{b}})}\BibitemShut
  {NoStop}%
\bibitem [{\citenamefont {Wallentowitz}\ and\ \citenamefont
  {Vogel}(1996)}]{wallentowitz96}%
  \BibitemOpen
  \bibfield  {author} {\bibinfo {author} {\bibfnamefont {S.}~\bibnamefont
  {Wallentowitz}}\ and\ \bibinfo {author} {\bibfnamefont {W.}~\bibnamefont
  {Vogel}},\ }\bibfield  {title} {\bibinfo {title} {Unbalanced homodyning for
  quantum state measurements},\ }\href
  {https://doi.org/10.1103/PhysRevA.53.4528} {\bibfield  {journal} {\bibinfo
  {journal} {Phys. Rev. A}\ }\textbf {\bibinfo {volume} {53}},\ \bibinfo
  {pages} {4528} (\bibinfo {year} {1996})}\BibitemShut {NoStop}%
\bibitem [{\citenamefont {Cahill}\ and\ \citenamefont
  {Glauber}(1969{\natexlab{a}})}]{cahill69}%
  \BibitemOpen
  \bibfield  {author} {\bibinfo {author} {\bibfnamefont {K.~E.}\ \bibnamefont
  {Cahill}}\ and\ \bibinfo {author} {\bibfnamefont {R.~J.}\ \bibnamefont
  {Glauber}},\ }\bibfield  {title} {\bibinfo {title} {Density operators and
  quasiprobability distributions},\ }\href
  {https://doi.org/10.1103/PhysRev.177.1882} {\bibfield  {journal} {\bibinfo
  {journal} {Phys. Rev.}\ }\textbf {\bibinfo {volume} {177}},\ \bibinfo {pages}
  {1882} (\bibinfo {year} {1969}{\natexlab{a}})}\BibitemShut {NoStop}%
\bibitem [{\citenamefont {Cahill}\ and\ \citenamefont
  {Glauber}(1969{\natexlab{b}})}]{cahill69a}%
  \BibitemOpen
  \bibfield  {author} {\bibinfo {author} {\bibfnamefont {K.~E.}\ \bibnamefont
  {Cahill}}\ and\ \bibinfo {author} {\bibfnamefont {R.~J.}\ \bibnamefont
  {Glauber}},\ }\bibfield  {title} {\bibinfo {title} {Ordered expansions in
  boson amplitude operators},\ }\href
  {https://doi.org/10.1103/PhysRev.177.1857} {\bibfield  {journal} {\bibinfo
  {journal} {Phys. Rev.}\ }\textbf {\bibinfo {volume} {177}},\ \bibinfo {pages}
  {1857} (\bibinfo {year} {1969}{\natexlab{b}})}\BibitemShut {NoStop}%
\bibitem [{\citenamefont {Innocenti}\ \emph {et~al.}(2022)\citenamefont
  {Innocenti}, \citenamefont {Lachman},\ and\ \citenamefont
  {Filip}}]{Innocenti2022}%
  \BibitemOpen
  \bibfield  {author} {\bibinfo {author} {\bibfnamefont {L.}~\bibnamefont
  {Innocenti}}, \bibinfo {author} {\bibfnamefont {L.}~\bibnamefont {Lachman}},\
  and\ \bibinfo {author} {\bibfnamefont {R.}~\bibnamefont {Filip}},\ }\bibfield
   {title} {\bibinfo {title} {Nonclassicality detection from few {Fock}-state
  probabilities},\ }\href {https://doi.org/10.1038/s41534-022-00538-y}
  {\bibfield  {journal} {\bibinfo  {journal} {npj Quantum Inf.}\ }\textbf
  {\bibinfo {volume} {8}},\ \bibinfo {pages} {30} (\bibinfo {year}
  {2022})}\BibitemShut {NoStop}%
\bibitem [{\citenamefont {Kovtoniuk}\ \emph {et~al.}(2022)\citenamefont
  {Kovtoniuk}, \citenamefont {Yeremenko}, \citenamefont {Ryl}, \citenamefont
  {Vogel},\ and\ \citenamefont {Semenov}}]{Kovtoniuk2022}%
  \BibitemOpen
  \bibfield  {author} {\bibinfo {author} {\bibfnamefont {V.~S.}\ \bibnamefont
  {Kovtoniuk}}, \bibinfo {author} {\bibfnamefont {I.~S.}\ \bibnamefont
  {Yeremenko}}, \bibinfo {author} {\bibfnamefont {S.}~\bibnamefont {Ryl}},
  \bibinfo {author} {\bibfnamefont {W.}~\bibnamefont {Vogel}},\ and\ \bibinfo
  {author} {\bibfnamefont {A.~A.}\ \bibnamefont {Semenov}},\ }\bibfield
  {title} {\bibinfo {title} {Nonclassical correlations of radiation in relation
  to {Bell} nonlocality},\ }\href {https://doi.org/10.1103/PhysRevA.105.063722}
  {\bibfield  {journal} {\bibinfo  {journal} {Phys. Rev. A}\ }\textbf {\bibinfo
  {volume} {105}},\ \bibinfo {pages} {063722} (\bibinfo {year}
  {2022})}\BibitemShut {NoStop}%
\bibitem [{\citenamefont {Boyd}\ and\ \citenamefont
  {Vandenberghe}(2004)}]{boyd_book}%
  \BibitemOpen
  \bibfield  {author} {\bibinfo {author} {\bibfnamefont {S.}~\bibnamefont
  {Boyd}}\ and\ \bibinfo {author} {\bibfnamefont {L.}~\bibnamefont
  {Vandenberghe}},\ }\href@noop {} {\emph {\bibinfo {title} {Convex
  Optimization}}}\ (\bibinfo  {publisher} {Cambridge University Press},\
  \bibinfo {address} {Cambridge, England},\ \bibinfo {year} {2004})\BibitemShut
  {NoStop}%
\bibitem [{\citenamefont {Brunner}\ \emph {et~al.}(2014)\citenamefont
  {Brunner}, \citenamefont {Cavalcanti}, \citenamefont {Pironio}, \citenamefont
  {Scarani},\ and\ \citenamefont {Wehner}}]{Brunner2014}%
  \BibitemOpen
  \bibfield  {author} {\bibinfo {author} {\bibfnamefont {N.}~\bibnamefont
  {Brunner}}, \bibinfo {author} {\bibfnamefont {D.}~\bibnamefont {Cavalcanti}},
  \bibinfo {author} {\bibfnamefont {S.}~\bibnamefont {Pironio}}, \bibinfo
  {author} {\bibfnamefont {V.}~\bibnamefont {Scarani}},\ and\ \bibinfo {author}
  {\bibfnamefont {S.}~\bibnamefont {Wehner}},\ }\bibfield  {title} {\bibinfo
  {title} {Bell nonlocality},\ }\href
  {https://doi.org/10.1103/RevModPhys.86.419} {\bibfield  {journal} {\bibinfo
  {journal} {Rev. Mod. Phys.}\ }\textbf {\bibinfo {volume} {86}},\ \bibinfo
  {pages} {419} (\bibinfo {year} {2014})}\BibitemShut {NoStop}%
\bibitem [{\citenamefont {Horodecki}\ \emph {et~al.}(1996)\citenamefont
  {Horodecki}, \citenamefont {Horodecki},\ and\ \citenamefont
  {Horodecki}}]{horodecki96}%
  \BibitemOpen
  \bibfield  {author} {\bibinfo {author} {\bibfnamefont {M.}~\bibnamefont
  {Horodecki}}, \bibinfo {author} {\bibfnamefont {P.}~\bibnamefont
  {Horodecki}},\ and\ \bibinfo {author} {\bibfnamefont {R.}~\bibnamefont
  {Horodecki}},\ }\bibfield  {title} {\bibinfo {title} {Separability of mixed
  states: necessary and sufficient conditions},\ }\href
  {https://doi.org/http://dx.doi.org/10.1016/S0375-9601(96)00706-2} {\bibfield
  {journal} {\bibinfo  {journal} {Phys. Lett. A}\ }\textbf {\bibinfo {volume}
  {223}},\ \bibinfo {pages} {1 } (\bibinfo {year} {1996})}\BibitemShut
  {NoStop}%
\bibitem [{\citenamefont {Terhal}(2000)}]{terhal99}%
  \BibitemOpen
  \bibfield  {author} {\bibinfo {author} {\bibfnamefont {B.~M.}\ \bibnamefont
  {Terhal}},\ }\bibfield  {title} {\bibinfo {title} {Bell inequalities and the
  separability criterion},\ }\href
  {https://doi.org/https://doi.org/10.1016/S0375-9601(00)00401-1} {\bibfield
  {journal} {\bibinfo  {journal} {Phys. Lett. A}\ }\textbf {\bibinfo {volume}
  {271}},\ \bibinfo {pages} {319} (\bibinfo {year} {2000})}\BibitemShut
  {NoStop}%
\bibitem [{\citenamefont {Horodecki}\ \emph {et~al.}(2009)\citenamefont
  {Horodecki}, \citenamefont {Horodecki}, \citenamefont {Horodecki},\ and\
  \citenamefont {Horodecki}}]{horodecki09}%
  \BibitemOpen
  \bibfield  {author} {\bibinfo {author} {\bibfnamefont {R.}~\bibnamefont
  {Horodecki}}, \bibinfo {author} {\bibfnamefont {P.}~\bibnamefont
  {Horodecki}}, \bibinfo {author} {\bibfnamefont {M.}~\bibnamefont
  {Horodecki}},\ and\ \bibinfo {author} {\bibfnamefont {K.}~\bibnamefont
  {Horodecki}},\ }\bibfield  {title} {\bibinfo {title} {Quantum entanglement},\
  }\href {https://doi.org/10.1103/RevModPhys.81.865} {\bibfield  {journal}
  {\bibinfo  {journal} {Rev. Mod. Phys.}\ }\textbf {\bibinfo {volume} {81}},\
  \bibinfo {pages} {865} (\bibinfo {year} {2009})}\BibitemShut {NoStop}%
\bibitem [{\citenamefont {Rockafellar}(1970)}]{Rockafellar_book}%
  \BibitemOpen
  \bibfield  {author} {\bibinfo {author} {\bibfnamefont {R.~T.}\ \bibnamefont
  {Rockafellar}},\ }\href@noop {} {\emph {\bibinfo {title} {Convex Analysis}}}\
  (\bibinfo  {publisher} {Princeton University Press},\ \bibinfo {address}
  {Princeton},\ \bibinfo {year} {1970})\BibitemShut {NoStop}%
\bibitem [{\citenamefont {Hug}\ and\ \citenamefont {Weil}(2020)}]{Hug_book}%
  \BibitemOpen
  \bibfield  {author} {\bibinfo {author} {\bibfnamefont {D.}~\bibnamefont
  {Hug}}\ and\ \bibinfo {author} {\bibfnamefont {W.}~\bibnamefont {Weil}},\
  }\href@noop {} {\emph {\bibinfo {title} {Lectures on Convex Geometry}}}\
  (\bibinfo  {publisher} {Springer Cham},\ \bibinfo {year} {2020})\BibitemShut
  {NoStop}%
\bibitem [{\citenamefont {Filip}\ and\ \citenamefont
  {Lachman}(2013)}]{Filip2013}%
  \BibitemOpen
  \bibfield  {author} {\bibinfo {author} {\bibfnamefont {R.}~\bibnamefont
  {Filip}}\ and\ \bibinfo {author} {\bibfnamefont {L.}~\bibnamefont
  {Lachman}},\ }\bibfield  {title} {\bibinfo {title} {Hierarchy of feasible
  nonclassicality criteria for sources of photons},\ }\href
  {https://doi.org/10.1103/PhysRevA.88.043827} {\bibfield  {journal} {\bibinfo
  {journal} {Phys. Rev. A}\ }\textbf {\bibinfo {volume} {88}},\ \bibinfo
  {pages} {043827} (\bibinfo {year} {2013})}\BibitemShut {NoStop}%
\bibitem [{\citenamefont {Kelley}\ and\ \citenamefont
  {Kleiner}(1964)}]{kelley64}%
  \BibitemOpen
  \bibfield  {author} {\bibinfo {author} {\bibfnamefont {P.~L.}\ \bibnamefont
  {Kelley}}\ and\ \bibinfo {author} {\bibfnamefont {W.~H.}\ \bibnamefont
  {Kleiner}},\ }\bibfield  {title} {\bibinfo {title} {Theory of electromagnetic
  field measurement and photoelectron counting},\ }\href
  {https://doi.org/10.1103/PhysRev.136.A316} {\bibfield  {journal} {\bibinfo
  {journal} {Phys. Rev.}\ }\textbf {\bibinfo {volume} {136}},\ \bibinfo {pages}
  {A316} (\bibinfo {year} {1964})}\BibitemShut {NoStop}%
\bibitem [{\citenamefont {Uzunova}\ and\ \citenamefont
  {Semenov}(2022)}]{Uzunova2022}%
  \BibitemOpen
  \bibfield  {author} {\bibinfo {author} {\bibfnamefont {V.~A.}\ \bibnamefont
  {Uzunova}}\ and\ \bibinfo {author} {\bibfnamefont {A.~A.}\ \bibnamefont
  {Semenov}},\ }\bibfield  {title} {\bibinfo {title} {Photocounting statistics
  of superconducting nanowire single-photon detectors},\ }\href
  {https://doi.org/10.1103/PhysRevA.105.063716} {\bibfield  {journal} {\bibinfo
   {journal} {Phys. Rev. A}\ }\textbf {\bibinfo {volume} {105}},\ \bibinfo
  {pages} {063716} (\bibinfo {year} {2022})}\BibitemShut {NoStop}%
\bibitem [{\citenamefont {Semenov}\ \emph {et~al.}(2024)\citenamefont
  {Semenov}, \citenamefont {Samelin}, \citenamefont {Boldt}, \citenamefont
  {Sch\"unemann}, \citenamefont {Reiher}, \citenamefont {Vogel},\ and\
  \citenamefont {Hage}}]{Semenov2024}%
  \BibitemOpen
  \bibfield  {author} {\bibinfo {author} {\bibfnamefont {A.~A.}\ \bibnamefont
  {Semenov}}, \bibinfo {author} {\bibfnamefont {J.}~\bibnamefont {Samelin}},
  \bibinfo {author} {\bibfnamefont {C.}~\bibnamefont {Boldt}}, \bibinfo
  {author} {\bibfnamefont {M.}~\bibnamefont {Sch\"unemann}}, \bibinfo {author}
  {\bibfnamefont {C.}~\bibnamefont {Reiher}}, \bibinfo {author} {\bibfnamefont
  {W.}~\bibnamefont {Vogel}},\ and\ \bibinfo {author} {\bibfnamefont
  {B.}~\bibnamefont {Hage}},\ }\bibfield  {title} {\bibinfo {title}
  {Photocounting measurements with dead time and afterpulses in the
  continuous-wave regime},\ }\href
  {https://doi.org/10.1103/PhysRevA.109.013701} {\bibfield  {journal} {\bibinfo
   {journal} {Phys. Rev. A}\ }\textbf {\bibinfo {volume} {109}},\ \bibinfo
  {pages} {013701} (\bibinfo {year} {2024})}\BibitemShut {NoStop}%
\bibitem [{\citenamefont {Stolyarov}\ \emph {et~al.}(2023)\citenamefont
  {Stolyarov}, \citenamefont {Kliushnichenko}, \citenamefont {Kovtoniuk},\ and\
  \citenamefont {Semenov}}]{Stolyarov2023}%
  \BibitemOpen
  \bibfield  {author} {\bibinfo {author} {\bibfnamefont {E.~V.}\ \bibnamefont
  {Stolyarov}}, \bibinfo {author} {\bibfnamefont {O.~V.}\ \bibnamefont
  {Kliushnichenko}}, \bibinfo {author} {\bibfnamefont {V.~S.}\ \bibnamefont
  {Kovtoniuk}},\ and\ \bibinfo {author} {\bibfnamefont {A.~A.}\ \bibnamefont
  {Semenov}},\ }\bibfield  {title} {\bibinfo {title} {Photon-number resolution
  with microwave {J}osephson photomultipliers},\ }\href
  {https://doi.org/10.1103/PhysRevA.108.063710} {\bibfield  {journal} {\bibinfo
   {journal} {Phys. Rev. A}\ }\textbf {\bibinfo {volume} {108}},\ \bibinfo
  {pages} {063710} (\bibinfo {year} {2023})}\BibitemShut {NoStop}%
\bibitem [{\citenamefont {de~Dios~Pont}\ \emph {et~al.}(2023)\citenamefont
  {de~Dios~Pont}, \citenamefont {Ivanisvili},\ and\ \citenamefont
  {Madrid}}]{Pont2023}%
  \BibitemOpen
  \bibfield  {author} {\bibinfo {author} {\bibfnamefont {J.}~\bibnamefont
  {de~Dios~Pont}}, \bibinfo {author} {\bibfnamefont {P.}~\bibnamefont
  {Ivanisvili}},\ and\ \bibinfo {author} {\bibfnamefont {J.}~\bibnamefont
  {Madrid}},\ }\href@noop {} {\bibinfo {title} {A new proof of the description
  of the convex hull of space curves with totally positive torsion}} (\bibinfo
  {year} {2023}),\ \Eprint {https://arxiv.org/abs/2201.12932} {arXiv:2201.12932
  [math.PR]} \BibitemShut {NoStop}%
\bibitem [{\citenamefont {Frankel}(2012)}]{Frankel_book}%
  \BibitemOpen
  \bibfield  {author} {\bibinfo {author} {\bibfnamefont {T.}~\bibnamefont
  {Frankel}},\ }\href@noop {} {\emph {\bibinfo {title} {The geometry of
  physics: an introduction}}},\ \bibinfo {edition} {3rd}\ ed.\ (\bibinfo
  {publisher} {Cambridge University Press},\ \bibinfo {address} {Cambridge,
  England},\ \bibinfo {year} {2012})\BibitemShut {NoStop}%
\bibitem [{\citenamefont {Banaszek}\ \emph {et~al.}(2002)\citenamefont
  {Banaszek}, \citenamefont {Dragan}, \citenamefont {W\'odkiewicz},\ and\
  \citenamefont {Radzewicz}}]{banaszek2002}%
  \BibitemOpen
  \bibfield  {author} {\bibinfo {author} {\bibfnamefont {K.}~\bibnamefont
  {Banaszek}}, \bibinfo {author} {\bibfnamefont {A.}~\bibnamefont {Dragan}},
  \bibinfo {author} {\bibfnamefont {K.}~\bibnamefont {W\'odkiewicz}},\ and\
  \bibinfo {author} {\bibfnamefont {C.}~\bibnamefont {Radzewicz}},\ }\bibfield
  {title} {\bibinfo {title} {Direct measurement of optical quasidistribution
  functions: {M}ultimode theory and homodyne tests of {Bell}'s inequalities},\
  }\href {https://doi.org/10.1103/PhysRevA.66.043803} {\bibfield  {journal}
  {\bibinfo  {journal} {Phys. Rev. A}\ }\textbf {\bibinfo {volume} {66}},\
  \bibinfo {pages} {043803} (\bibinfo {year} {2002})}\BibitemShut {NoStop}%
\bibitem [{\citenamefont {Fiur\'a\ifmmode~\check{s}\else
  \v{s}\fi{}ek}(2024)}]{Fiurasek2024}%
  \BibitemOpen
  \bibfield  {author} {\bibinfo {author} {\bibfnamefont {J.}~\bibnamefont
  {Fiur\'a\ifmmode~\check{s}\else \v{s}\fi{}ek}},\ }\bibfield  {title}
  {\bibinfo {title} {Tight nonclassicality criteria for unbalanced hanbury
  brown--twiss measurement scheme with click detectors},\ }\href
  {https://doi.org/10.1103/PhysRevA.109.033713} {\bibfield  {journal} {\bibinfo
   {journal} {Phys. Rev. A}\ }\textbf {\bibinfo {volume} {109}},\ \bibinfo
  {pages} {033713} (\bibinfo {year} {2024})}\BibitemShut {NoStop}%
\bibitem [{\citenamefont {Farkas}(1902)}]{Farkas1902}%
  \BibitemOpen
  \bibfield  {author} {\bibinfo {author} {\bibfnamefont {J.}~\bibnamefont
  {Farkas}},\ }\bibfield  {title} {\bibinfo {title} {Theorie der einfachen
  {U}ngleichungen},\ }\href {https://doi.org/doi:10.1515/crll.1902.124.1}
  {\bibfield  {journal} {\bibinfo  {journal} {J. die Reine und Angew. Math.
  (Crelles Journal)}\ }\textbf {\bibinfo {volume} {1902}},\ \bibinfo {pages}
  {1} (\bibinfo {year} {1902})}\BibitemShut {NoStop}%
\end{thebibliography}%

\end{document}